\documentstyle[onecolumn,epsfig,float]{mn}
\newcommand{\beq}{\begin{equation}}
\newcommand{\eeq}{\end{equation}}
\newcommand{\beqar}{\begin{eqnarray}}
\newcommand{\eeqar}{\end{eqnarray}}
\newcommand{\beqars}{\begin{eqnarray*}}
\newcommand{\eeqars}{\end{eqnarray*}}
\newcommand{\bc}{\begin{center}}
\newcommand{\ec}{\end{center}}
\newcommand{\ben}{\begin{enumerate}}
\newcommand{\een}{\end{enumerate}}
\newcommand{\bit}{\begin{itemize}}
\newcommand{\eit}{\end{itemize}}
\def \cLa{{\cal L}_1^{\pm 1}}
\def \cLb{{\cal L}_2^{\pm 1}}
\def \cLc{{\cal L}_3^{\pm 1}}

\newcommand{\I}{\mbox{\rm i}}

\def \eps{\epsilon}
\def \veps{\varepsilon}
\def \d{\partial}
\def \lam{\lambda}
\def \non{\nonumber}
\renewcommand{\(}{\left(}
\renewcommand{\)}{\right)}

\title[Inertial modes of slowly rotating relativistic stars in the
Cowling approximation]{Inertial modes of slowly rotating
  relativistic stars in the Cowling approximation}

\author[J.~Ruoff, A.~Stavridis, K.D.~Kokkotas]
{J.~Ruoff, A.~Stavridis, K.~D.~Kokkotas  \\
  Department of Physics, Aristotle University of Thessaloniki,
  Thessaloniki 54006, Greece}

\date{Accepted ???? Month ??.
      Received 2002 Month ??;
      in original form 2002 Month ??}
\pagerange{\pageref{firstpage}--\pageref{lastpage}}
\pubyear{2002}
\begin{document}

\maketitle

\label{firstpage}

\begin{abstract}
  We study oscillations of slowly rotating relativistic barotropic as
  well as non-barotropic polytropic stars in the Cowling
  approximation, including first order rotational corrections. By
  taking into account the coupling between the polar and axial
  equations, we find that, in contrast to previous results, the $m=2$
  $r$ modes are essentially unaffected by the continuous spectrum and
  exist even for very relativistic stellar models.  We perform our
  calculations both in the time and frequency domain. In order to
  numerically solve the infinite system of coupled equations, we
  truncate it at some value $l_{\rm max}$.  Although the time
  dependent equations can be numerically evolved without any problems,
  the eigenvalue equations possess a singular structure, which is
  related to the existence of a continuous spectrum. This prevents the
  numerical computation of an eigenmode if its eigenfrequency falls
  inside the continuous spectrum. The properties of the latter depend
  strongly on the cut-off value $l_{\rm max}$ and it can consist of
  several either disconnected or overlapping patches, which are the
  broader the more relativistic the stellar model is. By discussing
  the dependence of the continuous spectrum as a function of both the
  cut-off value $l_{\rm max}$ and the compactness $M/R$, we
  demonstrate how it affects the inertial modes. Through the time
  evolutions we are able to show that some of the inertial modes can
  actually exist inside the continuous spectrum, but some cannot.  For
  more compact and therefore more relativistic stellar models, the
  width of the continuous spectrum strongly increases and as a
  consequence, some of the inertial modes, which exist in less
  relativistic stars, disappear.
\end{abstract}


\section{INTRODUCTION}

Non-radial oscillations of relativistic stars gained a lot of interest
in the last decades because of the possible detection of their
associated gravitational waves. Especially after the discovery that
the $r$ modes of rotating neutron stars are generically unstable to
gravitational radiation (Andersson 1998, Friedman \& Morsink 1998),
the interest in the rotational instabilities renewed.  Great effort
has been put into understanding this instability and its implications
on the spin evolution of both newly born and old accreting neutron
stars.  If not damped by viscous effects, the $r$ mode instability
should be a very efficient way to generate gravitational radiation.
For a recent account of the $r$-mode instability we refer to the
review by Andersson and Kokkotas (2001).

Up to now, most of the studies of the $r$-mode instability have been
done within the Newtonian framework. In a first attempt to study the
relativistic analogue, Kojima (1997, 1998) considered the axial
equations for slowly rotating stars in the so-called ``low-frequency
approximation''. He found a situation quite different from Newtonian
theory, in that to first order the relativistic frame dragging in
rotating stars gives rise to a continuous spectrum of frequencies
instead of a single $r$-mode frequency. In a subsequent work, Beyer \&
Kokkotas (1999) proved in a mathematically rigorous way the existence
of this continuous spectrum in the low-frequency approximation.

Lockitch, Andersson \& Friedman (2001) showed that for uniform density
stars, in addition to the continuous spectrum, Kojima's equations also
admits discrete mode solutions, the relativistic $r$ modes. Ruoff \&
Kokkotas (2001) and Yoshida (2001) extended this study to polytropic
and realistic models and showed that for certain stellar models, no
$r$ modes can be found, even when the low-frequency assumption is
dropped and the radiation reaction is included (Ruoff \& Kokkotas
2002; Yoshida \& Futamase 2001). In all these studies, however, only
the axial perturbations have been considered, and any coupling to the
polar equations has been neglected. Lockitch et al.~(2001) have argued
that this assumption is only justified for non-barotropic stars, as
the relativistic $r$ modes in barotropic stars should be hybrids,
i.e.~ they retain a non-vanishing polar contribution in the
non-rotating limit.

By including higher order rotational corrections and the coupling to
the polar equations, Kojima \& Hosonuma (2000) derived a fourth order
equation, an extension of Kojima's master equation, which as a
consequence does not have the singular structure of the latter.
Although not rigorously proved, it was argued that this could possibly
resolve the problem of the non-existence of the $r$ modes in certain
stellar models (Lockitch \& Andersson 2001).

Recently Yoshida \& Lee (2002) announced that they can find
relativistic $r$ modes in the Cowling approximation for non-barotropic
stellar models. They obtained their results by only including first
order rotational correction terms. However, they did not restrict
themselves to merely considering the coupling between the axial
equation of order $l$ to the polar equations of order $l+1$, but
instead they solved the complete system starting from $l=|m|$ up to
some value $l_{\rm max}$, which they chose sufficiently high enough to
ensure the convergence of the modes. Lockitch et al.~(2001) had
already formulated this problem in a somewhat different way, by
neglecting only the perturbations which vanish in the stationary
limit, such as the energy and pressure perturbations for instance, and
some of the metric variables. So far they have presented numerical
results only in a post Newtonian treatment of relativistic $r$ modes
for uniform density stars. Their numerical eigenfunctions were used by
Stergioulas \& Font (2001) as approximate $r$-mode initial data for a
3-D general-relativistic hydrodynamical time evolution in the Cowling
approximation. They found that the $r$ mode oscillates at a
predominately discrete frequency with an indication of the kinematical
drift, as predicted by Rezzolla, Lamb \& Shapiro (2000).

In this paper, we follow the approach of Yoshida \& Lee (2002) and
extend it also to the time domain. Furthermore, we consider both
barotropic and non-barotropic stellar models and discuss in detail how
the continuous spectrum appears and how it is depends on the cut-off
$l_{\rm max}$. By computing a sequence of stellar models with
different $M/R$ ratios, we study how the continuous spectrum affects
their inertial modes. In this paper, we call all the modes ``inertial
modes'', whose restoring force is mainly the Coriolis force. Thus we
consider the $r$ modes as a subclass of inertial modes whose motion
is dominated by the axial velocity field.

The paper is organized as follows: In Section 2 we present both the
time dependent and the eigenvalue forms of the coupled perturbation
equations. We first specialize on the axisymmetric case, where we
describe how to compute the continuous spectrum and show its
dependence on $l_{\rm max}$. The numerical methods and results are
presented in Section 3 and we conclude with the discussion in Section
4.  Throughout the paper we use geometrical units $G=c=1$.  The
over-dots and primes denote differentiation with respect to the time
and the radial coordinates, respectively. All perturbation variables
should carry indices $l$ and $m$, however, we conveniently omit them.
Only in cases, where the $l$ index is important, we will include it.
Finally, we assume $m \ge 0$, as one can always transform a $m < 0$
mode with frequency $\sigma$ into a $m > 0$ mode with frequency
$-\sigma$ by taking its complex conjugate.

\section{FORMULATION OF THE PROBLEM}

\subsection{Background model}

We consider a uniformly rotating relativistic star with angular
velocity $\Omega$, which is assumed to sufficiently small that any
deviation from sphericity can be neglected. The metric can then be
written as (Hartle 1967)
\begin{equation}
  \label{metric}
  ds^2 = -e^{2\nu}dt^2
  + e^{2\lambda} dr^2 + r^2\( d\theta^2  + \sin^2\theta d\phi^2\)
  - 2\omega\sin^2\theta dt d\phi.
\end{equation}
Assuming a perfect fluid star, the energy momentum tensor
\begin{equation}
  \label{en_mom_tensor}
  T_{\mu\nu} = \(p + \eps\) u_{\mu}u_{\nu} + pg_{\mu\nu},
\end{equation}
with pressure $p$, energy density $\eps$ and four-velocity components
\begin{eqnarray}
  u^t &=& e^{-\nu},\\
  u^{\phi} &=& \Omega e^{-\nu}.
\end{eqnarray}
We will use a relativistic polytropic equation of state given by
\begin{eqnarray}
  \label{eos_p}
  p &=& \kappa \rho^{\Gamma},\\
  \label{eos_eps}
  \eps &=& \rho + {p \over \Gamma -1},
\end{eqnarray}
where $\kappa$ denotes the polytropic constant, $\Gamma$ is the
polytropic index and $\rho$ the rest mass density. With this form of
the equation of state, the polytropic index $\Gamma$ is equivalent to
the adiabatic index and obeys the relation
\begin{equation}
  \Gamma = \frac{p + \eps}{p}\frac{dp}{d\eps}.
\end{equation}

\subsection{Perturbed equations of motion}

We assume the oscillations to be adiabatic, so that the relation
between the Eulerian pressure perturbation $\delta p$ and energy
density perturbation $\delta\eps$ is given by
\begin{equation}
  \label{adcond}
  \delta p = \frac{\Gamma_1p}{p + \eps}\delta\eps
  + p'\xi^r\(\frac{\Gamma_1}{\Gamma} - 1\),
\end{equation}
where $\Gamma_1$ represents the adiabatic index of the perturbed
configuration and $\xi^r$ is the radial component of the fluid
displacement vector $\xi^\mu$. The sound speed $C_s$ is
\begin{eqnarray}
  C^2_s &=& \frac{\Gamma_1}{\Gamma}\frac{dp}{d\eps}.
\end{eqnarray}
The complete set of the perturbed Einstein equations has been derived
using the BCL gauge (Battiston, Cazzola \& Lucaroni 1971) in Ruoff,
Stavridis \& Kokkotas (2001), which we will refer to as RSK. As we are
working in the Cowling approximation, we only need to consider the
five fluid perturbations, which are the three components of the
velocity perturbations $\delta u_i$, the enthalpy perturbation $H$ and
the radial component of the displacement vector $\xi^\mu$. Following
RSK, we expand these quantities as
\begin{eqnarray}
  \label{dur}
  \delta u_r &=& -e^{\nu} \sum_{l,m} u_1^{lm} Y_{lm},\\
  \label{duth}
  \delta u_{\theta} &=& -e^{\nu} \sum_{l,m} \left[
    \tilde u_2^{lm} \d_{\theta} Y_{lm}
    - \tilde u_3^{lm} {\d_{\phi} Y_{lm} \over \sin\theta } \right],\\
  \label{duphi}
  \delta u_{\phi} &=&  -e^{\nu}
  \sum_{l,m} \left[ \tilde u_2^{lm} \d_{\phi} Y_{lm}
    + \tilde u_3^{lm} \sin\theta\d_{\theta} Y_{lm} \right],\\
  \delta \eps &=& \sum_{l,m}{\frac{\(p + \eps\)^2}{\Gamma_1p}
    \(H^{lm} - \xi^{lm}\)Y_{lm}},\\
  \xi^r &=& \bigg[\nu'\(1 - \frac{\Gamma_1}{\Gamma}\)\bigg]^{-1}
  \sum_{l,m}{\xi^{lm} Y_{lm}}.
\end{eqnarray}
Later we will introduce new variables $u_2$ and $u_3$ instead of
$\tilde u_2$ and $\tilde u_3$. These include rotational corrections
and therefore differ from the definitions of $u_2$ and $u_3$ in RSK.
The time dependent perturbation equations follow directly from
Eqs.~(68)--(72) of RSK with all the metric perturbations set to zero:
\begin{eqnarray}
  \label{dtHa}
  \(\d_t + \I m\Omega\)H &=& C_s^2\bigg\{e^{2\nu-2\lam}\bigg[u_1' +
  \(2\nu' - \lam' + \frac{2}{r}\)u_1 - e^{2\lam}\frac{\Lambda}{r^2}
  \tilde u_2\bigg] + \I m\varpi H\bigg\} - \nu'e^{2\nu-2\lam}u_1,\\
  \label{dtu1a}
  \(\d_t + \I m\Omega\)u_1 &=& H'
  + \frac{p'}{\Gamma_1 p}\bigg[\(\frac{\Gamma_1}{\Gamma} - 1\)H + \xi\bigg]
  - \bigg[\omega' + 2\varpi\(\nu' - \frac{1}{r}\)\bigg]
  \(\I m\tilde u_2 + \cLa\tilde u_3\),\\
  \label{dtu2a}
  \(\d_t + \I m\Omega\)\tilde u_2 &=& H
  + \frac{2\varpi}{\Lambda}\(\I m\tilde u_2 + \cLc\tilde u_3\)
  - \frac{\I mr^2}{\Lambda}A,\\
  \label{dtu3a}
  \(\d_t + \I m\Omega\)\tilde u_3 &=&
  \frac{2\varpi}{\Lambda}\(\I m\tilde u_3 - \cLc\tilde u_2\)
  + \frac{r^2}{\Lambda}\cLb A,\\
  \label{dtxia}
  \(\d_t + \I m\Omega\)\xi &=& \nu'\(\frac{\Gamma_1}{\Gamma} - 1\)
  e^{2\nu-2\lam}u_1,
\end{eqnarray}
where
\begin{eqnarray}
  \label{A}
  A &=& \varpi C_s^2e^{-2\lam}\left[u_1' + \(2\nu' - \lam' + \frac{2}{r}\)u_1
    - e^{2\lam}\frac{\Lambda}{r^2}\tilde u_2\right]
  + \left[\varpi\(\nu' - \frac{2}{r}\) + \omega'\right]
  e^{-2\lam}u_1
\end{eqnarray}
and
\begin{eqnarray}
  \Lambda &=& l(l+1).
\end{eqnarray}
The operators $\cLa$, $\cLb$, and $\cLc$ are the same as in RSK
and are defined by their action on a perturbation variable $P^{lm}$
\begin{eqnarray}
  \label{cLa}
  \cLa P^{lm} &=& (l-1)Q_{lm}P^{l-1m} - (l+2)Q_{l+1m}P^{l+1m},\\
  \label{cLb}
  \cLb P^{lm} &=& -(l+1)Q_{lm}P^{l-1m} + lQ_{l+1m}P^{l+1m},\\
  \label{cLc}
  \cLc P^{lm} &=& (l-1)(l+1)Q_{lm}P^{l-1m} + l(l+2)Q_{l+1m}P^{l+1m},
\end{eqnarray}
with
\begin{eqnarray}
  \label{Q}
  Q_{lm} &:=& \sqrt{\frac{(l-m)(l+m)}{(2l-1)(2l+1)}}.
\end{eqnarray}
By changing to new variables, we can simplify the above set of
evolution equations. To this end we use Eq.~(\ref{dtHa}) and rewrite
Eq.~(\ref{A}) as
\begin{eqnarray}
  A &=& \varpi e^{-2\nu}\d_tH + e^{-2\lam}Bu_1 + O(\Omega^2)
\end{eqnarray}
with
\begin{eqnarray}
  B &=& \omega' + 2\varpi\(\nu' - \frac{1}{r}\).
\end{eqnarray}
Neglecting second order terms and moving the terms including $\d_tH$
to the lefthand side, we can write Eqs.~(\ref{dtu2a}) and
(\ref{dtu3a}) as
\begin{eqnarray}
  \label{u2t}
  \d_t\(\tilde u_2 + \frac{\I m\varpi r^2}{\Lambda}e^{-2\nu} H\)
  + \I m\Omega\tilde u_2 &=& H
  + 2\frac{\varpi}{\Lambda}\(\I m\tilde u_2 + \cLc\tilde u_3\)
  - \frac{\I mr^2}{\Lambda}e^{-2\lam}Bu_1,\\
  \label{u3t}
  \d_t\(\tilde u_3 - \frac{\varpi r^2}{\Lambda}e^{-2\nu}\cLb H\)
  + \I m\Omega\tilde u_3 &=&
  2\frac{\varpi}{\Lambda}\(\I m\tilde u_3 - \cLc\tilde u_2\)
  + \frac{r^2}{\Lambda}e^{-2\lam}B\cLb u_1.
\end{eqnarray}
This suggests to define new variables
\begin{eqnarray}
  u_2 &:=& \tilde u_2 + \I m\frac{\varpi r^2}{\Lambda}e^{-2\nu} H,\\
  u_3 &:=& \tilde u_3 - \frac{\varpi r^2}{\Lambda}e^{-2\nu}\cLb H,
\end{eqnarray}
and rewrite Eqs.~(\ref{dtHa})--(\ref{dtxia}) in terms of $u_2$ and $u_3$.
After discarding any second order terms introduced by this replacement
the perturbations equations now read
\begin{eqnarray}
  \label{dtHb}
  \(\d_t + \I m\Omega\)H &=& e^{2\nu-2\lam}\bigg\{C_s^2\bigg[u_1' +
  \(2\nu' - \lam' + \frac{2}{r}\)u_1 - e^{2\lam}\frac{\Lambda}{r^2}u_2
  + 2\I m\varpi e^{2\lam-2\nu}H\bigg] - \nu'u_1\bigg\},\\
  \label{dtu1b}
  \(\d_t + \I m\Omega\)u_1 &=& H'
  + \frac{p'}{\Gamma_1 p}\bigg[\(\frac{\Gamma_1}{\Gamma} - 1\)H + \xi\bigg]
  - B\(\I mu_2 + \cLa u_3\),\\
  \label{dtu2b}
  \(\d_t + \I m\Omega\)u_2 &=& H
  + 2\frac{\varpi}{\Lambda}\(\I mu_2 + \cLc u_3\)
  - \frac{\I mr^2}{\Lambda}e^{-2\lam}Bu_1,\\
  \label{dtu3b}
  \(\d_t + \I m\Omega\)u_3 &=&
  2\frac{\varpi}{\Lambda}\(\I m u_3 - \cLc u_2\)
  + \frac{r^2}{\Lambda}e^{-2\lam}B\cLb u_1,\\
  \label{dtxib}
  \(\d_t + \I m\Omega\)\xi &=& \nu'\(\frac{\Gamma_1}{\Gamma} - 1\)
  e^{2\nu-2\lam}u_1.
\end{eqnarray}
In order to find the eigenmodes of this coupled system of equations,
we assume our perturbations variables to have a harmonic time
dependence $\exp(\I\sigma t)$. By replacing all time derivatives by
$\I\sigma$ and letting $H \rightarrow \I H$, $u_3 \rightarrow \I u_3$
and $\xi \rightarrow \I \xi$ in order to obtain a purely real valued
system of equations, we obtain two ODEs for $H$ and $u_1$ and three
algebraic relations for $u_2$, $u_3$ and $\xi$. The relation for
$\xi$, which follows from Eq.~(\ref{dtxib}) is rather simple and can
be immediately used to eliminate $\xi$, giving
\begin{eqnarray}
  \label{drH}
  H' &=& \(\sigma + m\Omega\)u_1
  - \frac{p'}{\Gamma_1 p}\(\frac{\Gamma_1}{\Gamma} - 1\)\bigg[H
  -\(\sigma + m\Omega\)^{-1}\nu'e^{2\nu-2\lam}u_1\bigg]
  + B\(mu_2 + \cLa u_3\),\\
  \label{dru1}
  u_1' &=& -\(2\nu' - \lam' + \frac{2}{r}\)u_1 + 2 m\varpi e^{2\lam-2\nu}H
  + e^{2\lam}\frac{\Lambda}{r^2}u_2
  - C_s^{-2}\bigg[\(\sigma + m\Omega\)e^{2\lam-2\nu} H - \nu'u_1\bigg],\\
  \label{u2alg}
  u_2 &=& \Sigma^{-1}\bigg[H - \frac{mr^2}{\Lambda}e^{-2\lam} Bu_1
  + \frac{2\varpi}{\Lambda}\cLc u_3\bigg],\\
  \label{u3alg}
  u_3 &=& -\Sigma^{-1}\bigg[\frac{r^2}{\Lambda}e^{-2\lam}B\cLb u_1
  - \frac{2\varpi}{\Lambda}\cLc u_2\bigg],
\end{eqnarray}
where we have defined
\begin{eqnarray}
  \Sigma := \sigma + m\Omega - \frac{2m\varpi}{\Lambda}.
\end{eqnarray}
The relations (\ref{u2alg}) and (\ref{u3alg}) cannot be solved
directly for $u_2$ and $u_3$ as they involve the coupling operators
${\cal L}_i^{\pm 1}$. We should stress again that because of this
coupling, both sets of equations (\ref{dtHb})--(\ref{dtxib}) and
(\ref{drH})--(\ref{u3alg}) have to be considered as infinite systems
with $l$ running from $m$ to infinity. There is no coupling to
equations with $l < m$, since from its definition (\ref{Q}) it follows
that the coupling coefficient $Q_{mm}$, which would couple to $l=m-1$,
is zero.  Furthermore, the equations form two independent sets, each
belonging to a different parity.  In the non-rotating case, one
usually distinguishes between polar and axial perturbations. Under
parity transformation polar perturbations change sign as $(-1)^l$,
axial perturbations as $(-1)^{l+1}$. In the rotating case, however,
the polar and axial equations are coupled, and therefore the
distinction between polar and axial modes cannot be upheld any more.
However, the equations do not mix the overall parity as can be seen as
follows. A polar equation with even $l=m$ has even parity. It is
coupled to an axial equation with $l+1$, whose parity is also even.
The next coupling is again to a polar equation with $l+2$, thus having
even parity. As this continues in the same manner, this means that for
even $m$, the complete coupled system with a leading polar equation
has even parity.  Conversely, the other system starting with a leading
axial equation has odd parity. For odd $m$, we obtain the reversed
situation.  This implies that for any given $m$, it makes sense to
distinguish the modes according to their overall parity.  Lockitch \&
Friedman (2000) introduced the notion of polar or axial led modes,
depending on whether the leading equations with $l=m$ are axial or
polar.  For even $m$, the polar led modes have even parity and the
axial modes odd parity, and vice versa for odd $m$.  The axisymmetric
case $m=0$ is somewhat special in that only the even parity modes
start with $l=0$ whereas the odd parity modes have to start with
$l=1$, as there are no axial $l=0$ equations.  In the following
section, we will restrict ourselves to positive $m$.

\subsection{The Continuous Spectrum}

For simplicity, we now focus on axisymmetric perturbations with $m=0$,
which leads to considerable simplifications of the time independent
equations (\ref{drH})--(\ref{u3alg}). With the expansion of the
operators ${\cal L}^{\pm1}_i$ according their definitions (\ref{cLa})
-- (\ref{cLc}), the $m=0$ equations read (we now explicitly include
the index $l$):
\begin{eqnarray}
  \label{H}
  (H^l)' &=& \sigma u_1^l
  + A\(l(l-1)Q_{l}u_3^{l-1} - (l+1)(l+2)Q_{l+1}u_3^{l+1}\),\\
  \label{u1}
  (u_1^l)' &=& -\(2\nu' - \lam' + \frac{2}{r}\)u_1^l
  + e^{2\lam}\frac{\Lambda}{r^2}u_2^l
  - C_s^{-2}\bigg[\sigma e^{2\lam-2\nu} H^l - \nu'u_1^l\bigg],\\
  \label{u2}
  u_2^l &=& \sigma^{-1}\bigg[H^l
  + 2\varpi\((l-1)Q_{l}u_3^{l-1} + (l+2)Q_{l+1}u_3^{l+1}\)\bigg],\\
  \label{u3}
  u_3^l &=& \sigma^{-1}\bigg[
  r^2e^{-2\lam}A
  \(Q_lu_1^{l-1} - Q_{l+1}u_1^{l+1}\)
  + 2\varpi\((l-1)Q_{l}u_2^{l-1} + (l+2)Q_{l+1}u_2^{l+1}\)\bigg],
\end{eqnarray}
with
\begin{eqnarray}
  Q_{l} &:=& \(4l^2 - 1\)^{-1/2}.
\end{eqnarray}
For $l=0$, we have $u_2^0 = u_3^0 = 0$. To solve the ODEs (\ref{H})
and (\ref{u1}), we have to compute $u_2^l$ and $u_3^l$ in terms of
$H^l$ and $u_1^l$. This can be done the most easily by combining
$u_2^l$ and $u_3^l$ as well as $H^l$ and $u_1^l$ into vectors and
rewriting the two algebraic relations (\ref{u2}) and (\ref{u3}) as
\begin{eqnarray}
  \label{vect}
  \sigma\(\begin{array}{c}u_2^l\\u_3^l\end{array}\)
  &=& 2\varpi(l-1)Q_l\(
  \begin{array}{cc}0&1\\1&0\end{array}\)
  \(\begin{array}{c}u_2^{l-1}\\u_3^{l-1}\end{array}\)
  + 2\varpi(l+2)Q_{l+1}\(
  \begin{array}{cc}0&1\\1&0\end{array}\)
  \(\begin{array}{c}u_2^{l+1}\\u_3^{l+1}\end{array}\)\non\\
  &&{}+ r^2e^{-2\lam}AQ_l\(
  \begin{array}{cc}0&0\\0&1\end{array}\)
  \(\begin{array}{c}H^{l-1}\\u_1^{l-1}\end{array}\)
  + \(
  \begin{array}{cc}1&0\\0&0\end{array}\)
  \(\begin{array}{c}H^l\\u_1^l\end{array}\)
  + r^2e^{-2\lam}AQ_{l+1}\(
  \begin{array}{cc}0&0\\0&-1\end{array}\)
  \(\begin{array}{c}H^{l+1}\\u_1^{l+1}\end{array}\).
\end{eqnarray}
Defining
\begin{eqnarray}
  u^l &=& \(\begin{array}{c}u_2^l\\u_3^l\end{array}\),\qquad
  s^l \;=\; \(\begin{array}{c}H^l\\u_1^l\end{array}\),\\
  U^l_- &=& -2\varpi(l-1)Q_l\(
  \begin{array}{cc}0&1\\1&0\end{array}\),\qquad
  U_\sigma \;=\; \sigma\(\begin{array}{cc}1&0\\0&1\end{array}\),\qquad
  U^l_+ \;=\; -2\varpi(l+2)Q_{l+1}\(
  \begin{array}{cc}0&1\\1&0\end{array}\),\\
  S^l_- &=& r^2e^{-2\lam}AQ_l\(
  \begin{array}{cc}0&0\\0&1\end{array}\),\qquad
  S \;=\; \(
  \begin{array}{cc}1&0\\0&0\end{array}\),\qquad
  S^l_+ \;=\; r^2e^{-2\lam}AQ_{l+1}\(
  \begin{array}{cc}0&0\\0&-1\end{array}\),
\end{eqnarray}
we can cast Eq.~(\ref{vect}) into a more compact form
\begin{eqnarray}
  \label{US}
  U^l_-u^{l-1} + U_\sigma u^l + U^l_+u^{l+1} &=&
  S^l_-s^{l-1}+ Ss^l+ S^l_+s^{l+1},\qquad l = 1, \dots, \infty
\end{eqnarray}
This, again, can be viewed as a matrix equation for two infinitely
dimensional matrices {\sf U} and {\sf S} acting on the vectors
\begin{eqnarray}
  u &=& \(u^1, u^2, \dots, u^l, \dots\)^{\rm T},\\
  s &=& \(s^1, s^2, \dots, s^l, \dots\)^{\rm T},
\end{eqnarray}
whose respective elements are the 2-vectors $u^l$ and $s^l$. Both {\sf
  U} and {\sf S} are tridiagonal block matrices, with each block given
by the above $2\times2$ matrices. Explicitly, {\sf U} is given by
\begin{eqnarray}
  {\sf U} &=& \(\begin{array}{cccccccc}
    U_\sigma&U^1_+&0&\dots&{}&{}&{}&\\
    U^2_-&U_\sigma&U^2_+&0&\dots&{}&{}&\\
    0&U^3_-&U_\sigma&U^3_+&0&\dots&{}&\\
    {}&{}&\ddots&\ddots&\ddots&{}&{}&\\
    {}&\dots&0&U^l_-&U_\sigma&U^l_+&0&\dots\\
    {}&{}&{}&{}&\ddots&\ddots&\ddots&\\
  \end{array}\).
\end{eqnarray}
A similar structure holds for {\sf S}. Now we can write
Eq.~(\ref{US}) as
\begin{eqnarray}
  \label{USb}
  \sum_{l'}{\mbox{\sf U}^{ll'}u^{l'}}
  &=& \sum_{l'}{\mbox{\sf S}^{ll'}s^{l'}}\;, \qquad l = 1\dots\infty,
\end{eqnarray}
which can be solved for $u^l$ by multiplying both sides with ${\sf
  U}^{-1}$
\begin{eqnarray}
  \label{matrix}
  u^l &=& \sum_{l'l''}{\(\mbox{\sf U}^{-1}\)^{ll'}
    \mbox{\sf S}^{l'l''}s^{l''}}\;, \qquad l = 1\dots\infty.
\end{eqnarray}
It is important to note that the matrix {\sf U} is $r$-dependent since
its elements $U^l_\pm$ contain the function $\varpi$. This raises the
question whether one can assure that {\sf U} will be invertible for
any frequency $\sigma$ at each value of $r$ in the stellar interior.
For {\sf U} to be invertible its determinant must not vanish.

In the Newtonian limit $\varpi\rightarrow\Omega$, the matrix {\sf U}
does not depend on $r$ any more, and one can easily show that {\sf U}
becomes singular only for a discrete set of values of $\sigma$. These
frequencies represent the solutions of the homogeneous part ${\sf U}u
= 0$ of Eq.~(\ref{USb}). In the relativistic case, however, we have to
replace $\Omega$ by the $r$-dependent effective angular velocity
$\varpi$. Thus the zeroes of $\det {\sf U}$ now depend on the position
inside the star.  This means that each single Newtonian frequency will
be spread out in the relativistic case into a continuous band of
frequencies, determined by the values of $\varpi$ at the centre and at
the surface of the star, which we will denote by $\varpi_0$ and
$\varpi_R$, respectively. The total range of the continuous spectrum is
then the sum of all the individual bands originating from their
Newtonian discrete values.

In order to numerically perform the inversion and solve equation
(\ref{matrix}), we have to truncate the system at some value $l_{\rm max}
> 1$. Let us consider the simple case, where we truncate the system at
$l_{\rm max}=2$, i.e. we couple only $l=1$ with $l=2$ and neglect all $l
\ge 3$. Then the matrix {\sf U} is given by
\begin{eqnarray}
  {\sf U} &=& \(\begin{array}{cc}
    U_\sigma&U^1_+\\
    U^2_-&U_\sigma
  \end{array}\)
  \;=\;\(\begin{array}{cccc}
    \sigma&0&0&-6\varpi Q_2\\
    0&\sigma&-6\varpi Q_2&0\\
    0&-2\varpi Q_2&\sigma&0\\
    -2\varpi Q_2&0&0&\sigma
\end{array}\)
\end{eqnarray}
with $Q_2 = 1/\sqrt{15}$ and $\det \sf U = 0$ yields
\begin{eqnarray}
  \sigma^2 &=& \frac{4}{5}\,\varpi^2 \;=\; 0.8\,\varpi^2.
\end{eqnarray}
Hence, the range of the continuous spectrum is given by
\begin{eqnarray}
  0.894\,\varpi_0 \le \sigma_c \le 0.894\,\varpi_R,
\end{eqnarray}
if we restrict ourselves to positive frequencies. If we also include
the $l=3$ terms, we obtain
\begin{eqnarray}
  \sigma^2 &=& \frac{84}{49}\,\varpi^2 \;\approx\; 1.714\,\varpi^2.
\end{eqnarray}
and the range is shifted to
\begin{eqnarray}
  1.309\,\varpi_0 \le \sigma_c \le 1.309\,\varpi_R.
\end{eqnarray}
Going one step further and including $l=4$, we obtain two solutions
\begin{eqnarray}
  \sigma^2 &=& \frac{4}{441}\(147 \pm 42\sqrt{7}\)\varpi^2,
\end{eqnarray}
corresponding to the frequency intervals
\begin{eqnarray}
  1.530\,\varpi_0\le\sigma_c\le 1.530\,\varpi_R
\end{eqnarray}
for the plus sign and
\begin{eqnarray}
  0.5704\,\varpi_0\le\sigma_c\le 0.5704\,\varpi_R
\end{eqnarray}
for the minus sign. Adding $l=5$ will still yield two ranges and for
$l=6$ we obtain three. The pattern is as follows: When a coupling to a
higher even value of $l$ is included, one additional range of the
continuous spectrum will appear. Depending on the respective values of
$\varpi_0$ and $\varpi_R$, the various ranges might as well overlap.
This actually happens for the more relativistic stars since the
variation in $\varpi$ is much greater.

A similar picture also holds in the case $m\ne0$. Here, each coupling
to a higher value of $l$ results in one additional band of the
continuous spectrum. We have already stated that in the case of the
matrix {\sf U} becoming singular at some point inside the star, we
cannot invert it anymore and the integration of the eigenvalue system
(\ref{drH})--(\ref{u3alg}) fails. In the purely axial case, Ruoff \&
Kokkotas (2001) showed that by a series expansion one could still
obtain mathematically valid mode solutions inside the continuous
spectrum, however, only at the price of a having a divergence in the
associated fluid perturbation at the singular point. Thus they
concluded that these modes are unphysical and should be discarded. The
question is whether we can adopt this point of view and claim that
inside the continuous bands modes cannot exist.

It is clear that the widths of the continuous bands strongly depend on
the stellar parameters and go to zero in the Newtonian limit.  For
weakly relativistic stellar models, each individual band is fairly
small, but the total range, which is the sum of these small bands, can
become quite large. If it is true that modes should not exist inside
the continuous bands, then even for weakly relativistic stellar
models, there should be almost no true mode solutions. As we cannot
perform the integration of (\ref{drH})--(\ref{u3alg}) inside the
continuous bands, we cannot tell whether there still might be valid
mode solutions. Trying series expansion of the coupled equations
around the singular point will be quite involved. This is why we
rather rely on the evolution of the time dependent equations, which
are free of singularities. By examining the Fourier spectra we can
assess whether or not we can find modes inside the continuous
spectrum. As we shall show and discuss, there are indeed modes which
lie inside the continuous bands. However, as the stellar models become
more compact, the continuous spectrum has the ability to destroy more
and more of the inertial modes.

\section{Numerical methods and results}

The numerical evolution of the coupled equations does not present any
severe numerical problems. Although one might encounter long term
instabilities, if a large number of $l$ is included, these can usually
be overcome by increasing the spatial resolution or shifting the
numerical origin some grid points away from zero. Up to values of
about $l_{\rm max}=10$, evolutions with $200$ grid points inside the star
yield quite accurate results for the inertial modes. As the inertial
modes are located at at the lower end of the frequency spectrum, we
have to perform long time evolutions in order to obtain a sufficient
frequency resolution.  Usually, we have to evolve up to the order of
$500\,$ms, corresponding to a total number of time steps well above
$10^6$.

The numerical mode calculation is somewhat more troublesome. As
explained in the previous section, we cannot compute the modes once
they are inside the continuous spectrum, because the matrix {\sf U}
becomes singular and inversion is no longer possible. To find the
modes outside the continuous spectrum, we proceed as follows. We first
rescale the variables according to
\begin{eqnarray}
  H^l &\rightarrow& r^lH^l,\\
  u_1^l &\rightarrow& r^{l-1}u_1^l,\\
  u_2^l &\rightarrow& r^lu_2^l,\\
  u_3^l &\rightarrow& r^{l+1}u_3^l.
\end{eqnarray}
This ensures that all our new variables have Taylor expansions around
the origin starting with the constant term. To initiate the
integration from the centre toward the stellar surface, we have to
prescribe initial data for each $H^l$ and $u_1^l$. At each integration
step, we have to invert the matrix {\sf U} in order to compute $u_2^l$
and $u_3^l$. At the origin, $H^l$ and $u_1^l$ are not independent, but
are related to leading order by
\begin{eqnarray}\label{H0u1}
  H^l(0) &=& \(\sigma + m\Omega - \frac{2m\varpi_0}{l}
  - 4\varpi_0^2\Sigma_{l-1}^{-1}\frac{(2l+1)(l-1)}{l^2}Q^2_{lm}\)
  \frac{u_1^l(0)}{l}.
\end{eqnarray}
To leading order, however, there is no relation between $H^l$ and any
other $H^{l'}$, so that we have $n = l_{\rm max} - m + 1$ degrees of
freedom for the initial values, as $l$ runs from $m$ to $l_{\rm max}$.
The frequency $\sigma$ represents a further degree of freedom, which
we have to choose deliberately. The missing boundary conditions, which
fix these degrees of freedom, come from the requirement of a vanishing
Lagrangian pressure perturbation at the stellar surface.  Since this
requirement translates into the simultaneous vanishing of the
quantities
\begin{eqnarray}
 \Delta p^l &:=& \bigg[\(\sigma + m\Omega\)e^{2\lam-2\nu} H^l
  - \nu'u_1^l\bigg]_{r=R}, \quad m \le l \le l_{\rm max},
\end{eqnarray}
we obtain $l_{\rm max} - m + 1$ conditions. We have thus reduced the
number of degrees of freedom to one, which is the value one expects
for linear systems, as the overall amplitude of the perturbations
remains arbitrary. Let us denote a set of initial data for the $H^l$
by ${\cal H}_k := \(H^m_k, \dots, H^{l_{\rm max}}_k\)$. In order to
compute the modes, we choose a frequency $\sigma$ and perform $n$
integrations for the following $n$ different sets of initial data:
${\cal H}_1 = \(1, 0, \dots, 0\), {\cal H}_2 = \(0, 1, 0, \dots, 0\),
\dots, {\cal H}_n = \(0, \dots, 1\)$.  The respective initial data for
$u_1$ follow from (\ref{H0u1}).  For each set ${\cal H}_k$, we compute
the $n$ Langragian pressure perturbations $\Delta p^l_k$, which in
general will not be zero.  After having performed the $n$
integrations, we end up with $n\times n$ values for the Langrangian
pressure perturbations $\Delta p^l_k$, which can be arranged in a
$n\times n$ ``pressure matrix'' {\sf P}. As the $n$ solutions
pertaining to the $n$ sets of initial data ${\cal H}_k$ are linearly
independent, we can try to use them to construct another solution by a
suitable linear combination, which satisfies $\Delta p^l = 0$ for all
$l$ simultaneously.  In other words we have to find $n$ coefficients
$a_k$, such that the linear combinations $a_k \Delta p^l_k$ vanish for
all $l$. This is possible if and only if the determinant of the
pressure matrix {\sf P} vanishes. The algorithm for finding the
eigenmode frequency is then obvious. For each $\sigma$, we construct
the pressure matrix {\sf P} and compute its determinante. If zero, we
have found an eigenmode.

To also compute the eigenfunction, we have to solve the homogeneous
system of equations
\begin{eqnarray}
  \sum_{k=m}^{l_{\rm max}}{a_k\Delta p^l_k} = 0,
  \qquad l = m \dots l_{\rm max}
\end{eqnarray}
for the coefficients $a_k$, which can be easily done by a ``singular
value decomposition'' of {\sf P} (see e.g.~Press et al.~1992). The
eigenfunction can be obtained by the linear combination of the $n$
solutions corresponding to the $n$ sets of initial data ${\cal H}_k$,
with the appropriate weighing coefficients $a_k$.

Although this method is quite easy to implement, it suffers from the
drawback of breaking down when $l_{\rm max} \ge m + 7$. This failure
results from the insufficient prescription of the initial data, where
we should include higher order terms in the Taylor expansion, as the
leading orders are set to zero for all $l$ but one. For the purpose of
computing the inertial modes for small values of $l$, however, our
method is robust enough, and we obtain perfect agreement with the
frequencies obtained from the time evolutions. Different methods are
used by Lockitch et al.~(2001) and Yoshida \& Lee (2002).

We have checked the consistency of our codes by feeding the evolution
code with the eigenfunctions obtained from the eigenvalue code. As
expected the time evolution yields a standing wave with one single
frequency.  Arbitrary initial data excite the $f$ and $p$ modes
together with the inertial and in the non-barotropic case also the $g$
modes. The frequencies found by Fourier transforming the time
sequences agree with those found from the mode calculation within less
than one per cent. This difference depends only on the numerical
resolution of the time evolutions and tends to zero for increasing
resolution.

\subsection{The axisymmetric case}

Let us first assess the influence of the rotation on the $f$ and $p$
modes. We take the same stellar model as Font et al.~(2001), which is
a $\Gamma = 2$ polytrope with $\kappa = 217.86\,$km$^2$. (Note their
value $\kappa = 100$ is in dimensionless units where $G = c = M_\odot
= 1$). With a central rest mass density $\rho_0 =
7.914\times10^{14}$g/cm$^3$, the total mass is $M = 1.4\,M_\odot$ and
the radius is $R = 14.15\,$km.  The rotation rate is given by $\Omega =
4230\,$s$^{-1}$ corresponding to $\veps = \Omega/\Omega_K = 0.5223$,
where $\Omega_K = \sqrt{M/R^3}$. In the following we consider only
barotropic perturbations, i.e.~we use $\Gamma_1 = 2$, thus suppressing
the $g$ modes. We will study non-barotropic perturbations in the
non-axisymmetric case.

In Table 1, we show the change in frequencies of the (quasi)radial
$f^0$ and $p^0$ modes, when the coupling to higher values of $l$ is
included. As we are considering axisymmetric perturbations, there is
no $m$-splitting and rotational effects only come from the coupling of
the equations with different $l$. For $l_{\rm max} = 0$, there are no
rotational corrections, i.e.~the equations describe the non-rotating
case. For $l_{\rm max} = 1$, the polar $l=0$ equations couple to the
axial $l=1$ equation for $u_3$, and we find a shift in the frequencies
towards higher values. This shift is about 3.7 per cent for the $f^0$
mode, 1.3 per cent for the $p^0_1$ mode and even less for higher
modes. When further couplings to higher $l$ are included, the
additional frequency corrections become smaller and the frequencies
converge rapidly. Beyond $l_{\rm max} = 3$, the frequency corrections
are less than 0.01 per cent.  Hence, we see that the quasi-radial $f$
and $p$ modes are quite insensitive to rotation. The effects of
rotation on the axisymmetric modes have been investigated using the
fully nonlinear equations in the Cowling approximation by Font et
al.~(2001). However, a comparison between our results and theirs is
not very meaningful as they study rapidly rotating models, which
undergo quite large radius and mass corrections with respect to the
non-rotating case. In the perturbation formalism, these corrections
appear only at the $\Omega^2$ level, so that we cannot account for
them when only first order corrections are included.  That these
corrections are important can be seen from the fact that their
frequencies decrease with increasing rotation rate while ours actually
increase. By including some of the second order corrections we have
convinced ourselves that the frequencies get pushed down to smaller
values, which then agree much better with the ones of Font et
al.~(2001). Therefore, we compare our values in Table 1 with theirs
only in the non-rotating limit. For the first four radial modes, they
find the frequencies $2708\,$Hz, $4547\,$Hz, $6320\,$Hz and
$8153\,$Hz, which agree very well with the first row of Table 1.  For
$m \ne 0$, we find the well-known linear splitting of the $f$ and $p$
modes into a lower and a higher frequency part.

\begin{table}
\centering\caption{$f^0$ and $p^0$ mode frequencies
as functions of $l_{\rm max}$ for the $1.4M_\odot$ model described in the
text. The rotation rate is $\veps = 0.5223$. The frequencies are given
in Hz. The $f$ and lowest $p$ mode experience the strongest rotational
corrections. For a coupling up to $l_{\rm max} = 3$, the mode frequencies
have converged.}
\begin{tabular}{*{6}{c}}
\hline
$l_{\rm max}$ & $f^0$  & $p^0_1$ & $p^0_2$ & $p^0_3$& $p^0_4$\\[0.5ex]
\hline
0 & 2687 & 4551 & 6344 & 8111 & 9867\\[0.5ex]
1 & 2787 & 4610 & 6386 & 8144 & 9894\\[0.5ex]
2 & 2795 & 4613 & 6387 & 8145 & 9895\\[0.5ex]
3 & 2796 & 4613 & 6387 & 8145 & 9895\\[0.5ex]
\hline \label{p modes}
\end{tabular}
\end{table}

Let us now turn our attention towards the inertial modes. In the
axisymmetric case, the first inertial modes appear, when coupling to
$l=2$ is included. As described in the previous section, a continuous
band of frequencies appears at the same time. In the time evolution,
the modes can be distinguished from the continuous band as follows.
Place some observers at different locations inside the star.  Take the
FFT of each observer's time series and plot the various spectra on top
of each other. Those peaks that coincide for all observers belong to
modes, while those peaks which are different for each observer
correspond to frequencies from the continuous spectrum.

As discussed above, the location of the continuous spectrum depends
strongly on how many $l$ are taken into account. Depending on
$l_{\rm max}$, it can consist of several patches. The width of each patch
is determined by the compactness of the stellar models and shrinks to
a set of discrete values in the Newtonian limit. For very compact
stars, the patches can become quite broad and in general overlap. In
the following, we will investigate the behaviour of both the
continuous spectrum and the inertial modes as functions of the stellar
compactness $M/R$ and the value of $l_{\rm max}$. As a first step,
however, it is helpful to understand the behaviour of the inertial
modes for an almost Newtonian stellar model, where the ranges of the
continuous spectrum are very narrow.

Lindblom \& Ipser (1999) have shown that for the Newtonian Maclaurin
spheroids, the eigenvalue problem can be made separable by choosing
appropriate spheroidal coordinates. Hence, $l$ and $m$ are ``good''
quantum numbers, which can be used to label the modes. For each value
of $m > 0$ and $l > m$ there exist $l-m$ inertial modes. For $m = 0$,
one finds $l - 1$ mode solutions. In this case, however, the
eigenvalue problem can be written in terms of $\sigma^2$, hence both
$\sigma$ and $-\sigma$ are valid eigenfrequencies. For even $l$, there
is an odd number of eigenvalues, hence one eigenvalue has to be
$\sigma = 0$ (c.f.~Table 1 of Lindblom \& Ipser 1999). If one
restricts oneself to positive frequencies, there is no mode for $l=2$,
one mode for $l=3$ and 4, two modes for $l=5$ and 6 and so on.

In our case, $l$ is not a good quantum number any more, but we still
can assign a certain value $l$ to each mode as we shall demonstrate.
To this end we now study for an almost Newtonian stellar model with
compactness $M/R = 0.01$ how both the continuous spectrum and the
inertial modes are affected as $l_{\rm max}$ is successively
incremented.

Figure 1 shows the locations of the continuous bands (grey shaded
areas) together with the rotational modes one can find for different
values of $l_{\rm max}$, ranging from $l_{\rm max} = 2$ to 5. For
$l_{\rm max} = 2$, we have a single continuous band in whose vicinity
many modes can be found.  These become more densely spaced as they
approach the continuous spectrum. The modes to the left of the
continuous spectrum have even parity and the modes to the right have
odd parity. When we go to $l_{\rm max} = 3$, we obtain a similar
picture. The continuous spectrum has moved to the right and again is
surrounded by a large number of modes. In addition, we find a more
isolated mode closer to the position, where the continuous spectrum
for $l_{\rm max} = 2$ was located. This is actually the first ``true''
inertial mode, as it does not change if we further increment $l_{\rm
  max}$. All the other modes which we can find for $l_{\rm max} = 2$
and $l_{\rm max} = 3$ are ``fake'' modes, i.e.~they are artefacts due
to the truncation of the equations and dissappear again when the
coupling to higher $l$ is included. We will label the ``true''
inertial modes with $i^{l_0}_n$, where $l_0$ is the value of $l_{\rm
  max}$ for which it appears for the first time and does not change
its frequency significantly when $l_{\rm max}$ is further increased.
Since for a given $l_{\rm max}$, more than just one new inertial mode
can appear, we order these modes through the lower index $n$ with
ascending frequency, i.e. $\sigma^{l_0}_n < \sigma^{l_0}_{n+1}$.
According to this convention, the inertial mode appearing for $l_{\rm
  max} = 3$ is the $i^3_1$ mode, which has odd parity.  Our upper
index $l_0$ agrees with the definition of Yoshida \& Lee (2000a) and
differs from the one of Lockitch \& Friedman, who use $l_0-1$ to
denote the same mode. Yoshida \& Lee (2000a), however, use the lower
index $n$ in a somewhat different form, as they compute the modes in
the rotating frame, which then form a set with equal fractions of
negative and positive frequency modes. Yoshida \& Lee attach negative
indices $n$ to the negative frequencies modes and positive $n$ to the
positive frequency modes.

For $l_{\rm max} = 4$, the continuous spectrum consists of two patches,
and we find the $i^4_1$ mode, which now has even parity. For $l_{\rm max}
= 5$, these two patches shift to the right and two new inertial modes,
$i^5_1$ and $i^5_2$ appear. The emerging pattern is as follows: For
each even $l_{\rm max}$, a new patch of the continuous spectrum appears.
If $l_{\rm max} > 2$ is odd, $(l_{\rm max} - 1)/2$ new odd parity inertial
modes appear, if $l_{\rm max}$ is even, we find $(l_{\rm max} - 2)/2$ new even
parity modes. Besides the inertial modes for a given $l_{\rm max}$ appear
close to the positions where the patches of the continuous spectrum
were located for $l_{\rm max} - 1$.  Examination of the eigenfunctions
reveals that a mode $i^l_n$ has no node in the coefficient $H^l$, one
node in the coefficient $H^{l-2}$, and $n$ nodes in the coefficient
$H^{l-2n}$.

It can be noticed that for $l_{\rm max} = 4$ and 5, there are only
very few modes in the vicinity of the left patches of the continuous
spectrum, which is due to the fact that these patches are somewhat
broader than the other ones. If one considers less relativistic
models, the width of the patches shrink and more modes can be found
around them. In the Newtonian case, where all the patches degenerate
to single points, we expect an infinite number of modes around these
points. As the star becomes more relativistic and the continuous bands
expand, they will swallow more and more modes. In principle this is of
no relevance as these modes are not physical, anyway. The remaining
question is, whether or not the continuous spectra are able to swallow
the ``true'' inertial modes, if they grow broad enough.

The answer is depicted in Fig.~2, where we plot the inertial modes
together with the continuous spectra as functions of both the
compactness $M/R$ and $l_{\rm max}$. Here, we plot only the ``true''
inertial modes and not those resulting from the truncation of the
equations at $l_{\rm max}$ as in Fig.~1. The four panels illustrate
the situation for the different cut-off values $l_{\rm max}$ ranging
from 3 to 6. The compactness $M/R$ is from 0.01 (as in Fig.~1) to
0.22.  The stability limit with respect to radial collapse is reached
at $M/R = 0.215$. All stellar models have the same $\veps = 0.5223$,
which means that the less compact models rotate much more slowly than
the highly relativistic ones. Since the frequencies of the inertial
modes depend essentially linearly on the rotation rate, they would
become very small for less compact models. This is the reason why we
express them in units of $\Omega_K$. We thus see that the inertial
modes lie in the interval between 0 and 1 for the whole compactness
range.  (Actually they lie in the interval between -1 and 1, but as
negative and positive frequencies have the same magnitude, we plot
only the positive ones).

Starting with $l_{\rm max} = 3$, we have one single continuous
spectrum, which broadens considerably as the compactness $M/R$ is
increased. For small $M/R$, we also find the $i^3_1$ mode, which moves
to lower frequencies as $M/R$ increases.  However, the lower border of
the continuous spectrum decreases much faster and for $M/R \approx
0.17$ it reaches the mode.  For larger $M/R$ the mode cannot be found
any more. We stress again that once the mode would be inside the
continuous spectrum, the eigenvalue code breaks down because the
matrix $\sf U$ becomes singular and inversion is no longer possible.
In this case, we take the results of the time evolutions to decide
whether or not the mode exists inside the continuous spectrum. And
indeed, the time evolution shows that for $M/R > 0.17$ the mode does
not penetrate the continuous spectrum but instead disappears.

When we go to $l_{\rm max} = 4$, the same happens to the $i^4_1$ mode
which reaches the upper patch of the continuous spectrum at about $M/R
\approx 0.11$ and disappears for larger $M/R$. Surprisingly, the
$i^3_1$ mode now exists over the complete compactness range. Indeed,
the time evolution shows that it can even exist inside the continuous
spectrum for $M/R > 0.2$. This feature is even more apparent when the
coupling up to $l_{\rm max} = 5$ is included. In this case, the
$i^3_1$ mode is always inside the continuous spectrum over the whole
compactness range. However, the $i^4_1$ mode still does not penetrate
the continuous spectrum and the same is true for the $i^5_1$ and
$i^5_2$ modes. Notice also that for $M/R > 0.17$, the two patches of
the continuous spectrum begin to overlap.

For $l_{\rm max} =6$, both the $i^3_1$ and $i^3_1$ modes exist for the
complete compactness range. Again, the $i^5_1$ is always inside the
continuous spectrum, whereas the $i^5_1$ mode enters it only for $M/R
> 0.2$. The $i^5_2$ mode, in contrast, ceases to to exist for $M/R >
0.1$. For the $l_{\rm max} = 7$ case, which is not shown, only the
$i^3_1$, $i^5_1$ and $i^7_1$ modes exist for the complete compactness
range, most of the others exist only outside the continuous spectrum.
Some actually do penetrate the continuous spectrum, but if the star
becomes too compact, they eventually vanish.

The emerging picture is as follows. Only the modes with low
frequencies, such as the $i^3_1$, $i^5_1$ and $i^7_1$ modes for
instance are able to exist for the complete compactness range and even
inside the continuous spectrum. Those with higher frequencies exist
only for a much smaller range.  This might be attributed to the fact
that the patches of the continuous spectrum are much broader for
larger frequencies.  As a consequence, the total number of inertial
modes decreases as the stellar models become more relativistic.
However, this leaves the question why some of the modes can survive
inside the continuous spectrum and some cannot. We conjecture that
this has to do with the value of $l_{\rm max}$, where the mode occurs
for the first time. If for a certain value of $l_{\rm max}$ it appears
well outside the continuous spectrum, then it will survive if $l_{\rm
  max}$ is further increased and the mode suddenly finds itself inside
the continuous spectrum. In other words, it is not the mode which
moves inside the continuous spectrum as $l_{\rm max}$ is increased, it
is the continuous spectrum which shifts itself to the position where
the mode is located. If, however, for a certain $l_{\rm max}$ the
continuous spectrum is already sufficiently broad to occupy the
position where an inertial mode would appear, then this does not seem
to be able to establish itself, even when higher values of $l$ are
included.

\subsection{The non-axisymmetric case}

For $m \ne 0$, we find essentially the same picture as in the
axisymmetric case, i.e.~some of the modes can exist for the whole
compactness range while some cannot. Here, the most interesting
question, of course, is what happens to the relativistic $r$-modes and
in which way are they affected by the continuous spectrum.  One can
see that the relativistic case differs qualitatively from the
Newtonian one, as in the latter, one can immediately show that the
leading order of the $r$-mode frequency (in the inertial frame) is
given by
\begin{eqnarray}
  \label{r mode}
  \sigma &=& -m\Omega\(1 - \frac{2}{l(l+1)}\),
\end{eqnarray}
whereas the analogous relativistic calculation yields
\begin{eqnarray}
  \label{cs}
  \sigma(r) &=& -m\Omega\left[1 - \frac{2}{l(l+1)}
  \(1 - \frac{\omega(r)}{\Omega}\)\right].
\end{eqnarray}
Instead of a single frequency, we have a continuous spectrum of
frequencies, determined by the values of the frame dragging $\omega$
at the centre and surface of the star. Herein, we have chosen the sign
of $\sigma$ such that it yields negative frequencies for positive
values of $m$.

The above values for $\sigma$ were obtained by assuming that the $r$
mode is purely axial to leading order, and the coupling to the polar
equations is just a higher order correction. However, it was shown by
Lockitch et al.~(2001) that this is not true for barotropic
perturbations, as this assumption leads to some incompatibilities
within the polar equations. Instead, the barotropic $r$ mode
necessarily has to include polar contributions, which do not vanish in
the non-rotating limit. For a ``pure'' $r$ mode, the polar
contribution has to vanish. In the terminology of Lockitch et
al.~(2001), the relativistic barotropic $r$ mode is therefore an
``axial-led hybrid'' mode. In the non-barotropic case, things are
different, as the $r$ mode cannot have non-vanishing polar
contributions in the non-rotating limit. This is because for
non-rotating non-barotropic stellar models, there are no stationary
polar perturbations, since they would immediately excite $g$ modes.
For non-rotating barotropic stars on the other hand, the $g$ modes
form a degenerate space of zero-frequency modes, which is why the
axial inertial modes can have polar contributions.  Thus, for
non-barotropic stars, the $r$ mode should not be an axial-led hybrid,
but it should reduce in the non-rotating limit to a purely axial
stationary perturbation.  Nevertheless, we will call the inertial mode
$i_1^l$ with $l=m+1$ an $r$ mode, regardless whether the star is
barotropic or non-barotropic.  (Note that Lindblom \& Ipser call any
inertial mode a generalized $r$ mode).

Let us start with the barotropic case for $m=2$. As already mentioned
above, the axial equation for $l=2$ yields the band of frequencies
given by Eq.~(\ref{cs}). This is depicted in the left graph of Fig.~3,
where we show the continuous spectrum as a function of $M/R$. No
inertial modes are present. We have also included the Newtonian value
for the $r$ mode as given by Eq.~(\ref{r mode}). Note that this value
is completely independent of any stellar structure parameters. If we
now include the coupling to the polar equations with $l=3$, the
situation changes drastically. We now obtain two continuous frequency
bands and a mode which lies below the Newtonian $r$ mode value, the
further below the more compact the star is. Translating to positive
frequencies, this means that this mode always has a higher frequency
than the Newtonian $r$ mode. Inspection of the eigenfunction reveals
that the axial part fits nicely to a $r^{l+1}$ power law, i.e.
\begin{eqnarray}
  u_3^{l=2} &\sim& r^3,
\end{eqnarray}
which is exactly what one expects for the $r$ mode. Inclusion of
higher $l$ only gives minor frequency corrections, which tells us that
this is indeed a physical inertial mode, the relativistic version of
an $m = 2$ $r$ mode. According to our nomenclature for inertial
modes,the $r$ mode is actually an $i_1^3$ inertial mode. In Fig.~3, we
have also included the coupling up to $l_{\rm max} = 4$ and 5. Here,
each inclusion of one additional $l$ leads to another patch in the
continuous spectrum. Notice that the width of the uppermost and
lowermost patches is very broad, whereas the middle patches around
$\sigma/\Omega_L \approx -1$ remain very narrow, even for large $M/R$.
For each $l_{\rm max}$, we find $l_{\rm max}-m$ new inertial modes,
appearing very close to those frequencies which were occupied by the
continuous spectrum for $l_{\rm max}-1$. Again, some modes exist
throughout the whole compactness range, while some are destroyed by
the continuous spectrum above some $M/R$. In particular this is the
case for those modes residing at both ends of the frequency range, as
the continuous bands become very broad for large $M/R$.

To assess the accuracy of our results, we have compared our $r$ mode
frequencies with non-linear time evolutions. For the $1.4\,$M$_\odot$
neutron star model with $\Omega = 2180\,$s$^{-1}$, the non-linear
evolution in the Cowling approximation yields an $r$-mode frequency of
500$\,$Hz (Stergioulas, 2002). Our linear calculation gives a
frequency of $512\,$ Hz, which differs only by 2.5 per cent from the
non-linear value. For larger rotation rates the difference increases,
and for $\Omega = 4986\,$s$^{-1}$, the non-linear value is $1030\,$Hz,
whereas we find $1172\,$Hz, which is by 14 per cent off.

When turning to the non-barotropic case the picture becomes more
complicated because of the presence of the $g$ modes. When the star is
set into rotation, the $g$ and inertial modes start to interact.
Yoshida \& Lee (2000b) have studied in depth the behaviour of the $g$
and inertial modes as functions of the rotation rate and $\Gamma_1$
for Newtonian stars. In this paper, we would like to focus only on the
influence of the non-barotropicity on the $r$ mode.

In accordance with the results Yoshida \& Lee (2000b) we find that the
$r$ mode frequency remains almost unaffected as the star starts to
deviate from being barotropic. However, we find that for small
rotation rates, the eigenfunctions change in a quite drastic way if
$\Gamma_1$ starts to differ from $\Gamma$. The larger the rotation,
the more similar they get. We have already mentioned above that the
relativistic $r$ mode of a barotropic star has to be an axial-led
hybrid mode, which must retain its polar contribution in the
non-rotating limit, whereas for a non-barotropic star the polar
contribution must vanish. This can be readily checked by computing a
sequence of eigenfunctions for decreasing values of $\Omega$. These
results are shown in Fig.~4, were we contrast the $m=2$ $r$ modes for
a barotropic model with those of the corresponding non-barotropic
model with $\Gamma_1 = 3$. We only show the dominant parts of the
eigenfunctions, which are the axial contribution $u_3$ with $l=2$ and
the polar parts $H$, $u_1$ and $u_2$ with $l=3$. Furthermore, we
rescale the eigenfunctions in such a way that $u_3(R) = 1$.

For the barotropic sequence (left panels), we can see that the shape
of $u_3$ does not depend on the rotation rate and is always given by
$r^3$. The polar parts $H$, $u_1$ and $u_2$ actually do change along
the sequence. Whereas $H$ goes to zero, $u_1$ and $u_2$ approach some
finite values. This is in full agreement with Lockitch et al.~(2001),
who have shown that the non-trivial stationary perturbations must have
vanishing $H$ and non-vanishing $u_1$ and $u_2$. From Eq.~(\ref{dtHb})
is clear that for vanishing $H$ (and therefore vanishing $\dot H$),
$u_1$ and $u_2$ must satisfy
\begin{eqnarray}
   u_1' + \(2\nu' - \lam' + \frac{2}{r}\)u_1
   - e^{2\lam}\frac{l(l+1)}{r^2}u_2 &=& 0,
\end{eqnarray}
which is equivalent to Eq.~(3.23) of Lockitch et al.~(2001).

The non-barotropic sequence (right panels) is completely different.
For a large rotation rate $\veps = 0.2$, the eigenfunction is very
similar to the barotropic one. However, for smaller rotation rates,
they begin to deviate considerably. The shape of the axial part $u_3$
is no longer described by $r^3$, but now becomes zero throughout most
parts of the star and only peaks close to the surface. Our chosen
value of $\Gamma_1 = 3$ seems to be quite large, however, the same
effect can be observed for $\Gamma_1$ much close to 2, in this case,
it appear for much smaller rotation rates. In principle, we should not
expect this behaviour, instead the shape of $u_3$ should remain the
same and only the polar perturbations should go to zero. From Fig.~4
we can see that latter is actually the case and $H$ as well as $u_1$
and $u_2$ all go to zero in the non-rotating limit. We can track this
singular behaviour of $u_3$ back to the frame dragging $\omega$, for
if we set it to zero, thus simulating a Newtonian stellar model, the
eigenfunction of $u_3$ is well behaved; i.e.~it remains proportional
to $r^3$ for any rotation rate, whereas all the polar perturbations go
to zero in the non-rotating limit. We suppose that this singular
behaviour of $u_3$ is an artefact of the Cowling approximation, which
should be cured by taking into account the metric perturbations.

\section{Summary}

We have investigated the inertial modes of slowly rotating
relativistic stars in the Cowling approximation. Considering only
first order rotational corrections, we have performed our computations
in both the time and the frequency domain. Comparison with non-linear
results for the $f$, $p$ and $r$ modes showed very good agreement, at
least for slowly rotating stars. For larger rotation, second order
effects become important and would have to be included for accurate
mode calculations.

For the numerical calculations, the system of equations has to be
truncated at some cut-off value $l_{\rm max}$. This truncation,
however, introduces many spurious modes, which either vanish or shift
in frequencies when $l_{\rm max}$ is increased. The ``true'' inertial
modes, which appear for a given $l_{\rm max}$ are those which do not
change significantly when $l_{\rm max}$ becomes larger. For each $l >
m > 0$, one can find $l-m$ inertial modes, which can be labeled
$i^l_n$ ($1 \le n \le l-m$).  For the axisymmetric case $m=0$, there
are $l-1$ different modes, although in this case one mode has zero
frequencies and the others always come in pairs with positive and
negative frequencies.

We have shown that, as a result of the relativistic frame dragging,
there always exists a continuous spectrum, which, however, depends
very strongly on the number of coupled equations. For $l_{\rm max} =
2$, the continuous spectrum is confined to a single connected region,
with an increasing width for more relativistic stellar models. If
$l_{\rm max}$ is increased the continuous spectrum splits into an
increasing number of patches, which are disconnected for weakly
relativistic stars but tend to overlap for more compact stellar
models. For $m \ne 0$, each successive inclusion of a higher $l$
creates a new patch, whereas for $m=0$, only every second $l$ does so.
In the limit $l_{\rm max}\rightarrow\infty$ the continuous spectrum
should completely cover the frequency range of the inertial modes. In
previous work, which neglected the coupling between the equations, it
was stated that modes should not be able to exist inside the
continuous spectrum, as the eigenfunctions would become singular.
Through the time evolutions, however, we were able to show that there
are modes, which can exist inside the continuous spectrum.

We provide the following (phenomenological) explanation. If an
inertial mode can exist inside the continuous spectrum for a given
$l_{\rm max}$, then it had to appear for the first time outside the
continuous spectrum for some smaller $l_{\rm max}$. If a mode with
label $i^{l_0}_n$ appears outside the continuous spectrum for $l_{\rm
  max} = l_0$, then it will not be affected if the continuous spectrum
shifts itself on top of the mode for some $l_{\rm max} > l_0$. On the
other hand, if the continuous spectrum for $l_{\rm max} = l_0$ is
already so broad that it covers the position where the $i^{l_0}_n$
mode would appear, then this mode cannot come into existence.

Hence, it seems that for very relativistic stars, the continuous
spectrum is able to destroy some of the inertial modes. We should,
however, be still very cautious to draw any conclusions about the
(non)existence of these modes for a rapidly rotating neutron star. We
have seen that the continuous spectrum is very sensitive to the number
of equations that are coupled. As we are only taking into account the
first order rotational corrections, there is only coupling from $l$ to
$l\pm1$.  Inclusion of second order corrections leads to coupling to
$l\pm2$; however, this should not affect the continuous spectrum,
since in the relevant equations for $u_2$ and $u_3$, only terms
coupling to $H$ and $u_1$ arise. Third order corrections might do the
job and it could well be that they will affect the continuous spectrum
in such a way that modes which do not exist in the first order
analysis can now exist because the responsible patch could have moved
somewhere else.

So far, we have completely neglected any metric perturbations, as we
worked in the Cowling approximation. Lockitch et al.~(2001) included
some (non-radiative) metric perturbations, but restricted their
studies to a post Newtonian treatment.  Nevertheless it is clear that
we should expect some quantitative difference when including these
terms, i.e.~when we move from the Cowling approximation to the
low-frequency approximation.  Still, the continuous spectrum will
remain unaffected when metric variables are included, for it is only
the fluid equations which are responsible for its existence. By
including the metric perturbations, the modes might be affected in
such a way, however, that they get pushed out of the continuous
spectrum and we might find some of the modes which we cannot find in
the Cowling approximation. We think that this scenario seems to be
quite likely as the study of the purely axial $l=m$ case has shown.
There, the Cowling approximation leads to a purely continuous spectrum
without any mode solutions. In the low-frequency approximation, where
one retains a certain metric component, one is lead to Kojima's master
equation, which in certain cases admits $r$ mode solutions, whose
frequencies lie outside the continuous spectrum (Ruoff \& Kokkotas
2001; Yoshida 2001).  But the range of the continuous spectrum was
still the same as in the Cowling approximation and even when all the
metric perturbations were included (Ruoff \& Kokkotas 2002; Yoshida \&
Futamase 2001), the qualitative picture remained the same. Hence, we
conjecture that the inclusion of the metric perturbations in the fully
coupled equations will not affect the behaviour of the continuous
spectrum, but it certainly will shift the modes, and possibly in such
a way that some of them might be pushed outside the patches of the
continuous spectrum and therefore reappear.

As far as the $r$ modes are concerned, our results show that their
existence does not seem to called into question by the continuous
spectrum. The previous results by Ruoff \& Kokkotas (2001, 2002),
Yoshida (2001) and Yoshida \& Futamase (2002), which suggested that in
certain stellar models with rather soft equations of state, the
(purely axial) relativistic $r$ modes cannot exist due to the
existence of the continuous spectrum, seem to be artefacts arising
from neglecting the coupling to the polar equations. The inclusion of
this coupling shifts the location of the continuous spectrum and the
$r$ mode can exist well outside it for a large range of stellar
models. In this paper, we have only presented results for a single
polytropic equation of state with $\Gamma = 2$, but from our previous
results (Ruoff \& Kokkotas 2001), we know that for $\Gamma > 2$, the
width of the continuous spectrum actually shrinks and thus favours the
presence of inertial modes.  As most of the current realistic
equations of state are stiffer than a $\Gamma = 2$ polytrope, we
expect that the existence of $r$ modes should not be questioned. The
same should also hold for strange stars.

Nevertheless it is clear that the relativistic case is quite different
from the Newtonian one, and further studies are necessary to gain a
full understanding of the inertial modes of (rapidly) rotating
relativistic stars. In particular the metric perturbations have to be
included to obtain the correct $r$-mode frequencies and the associated
growth times. We cannot rule out that, in particular the latter will
deviate significantly from the results obtained from applying the
quadrupole formula to Newtonian stellar models.

\section*{ACKNOWLEDGMENTS}

We wish to thank Nikolaos Stergioulas and Ulrich Sperhake for useful
comments. J.R.~is supported by the Marie Curie Fellowship
No.~HPMF-CT-1999-00364.  A.S.~was supported by the Greek National
Scholarship foundation (I.K.Y.) during this work.This work has been
supported by the EU Programme 'Improving the Human Research Potential
and the Socio-Economic Knowledge Base' (Research Training Network
Contract HPRN-CT-2000-00137).

\label{lastpage}

\newpage

\begin{figure}
  \centering
  \vspace*{2cm}
  \epsfxsize=9cm
  \epsfbox{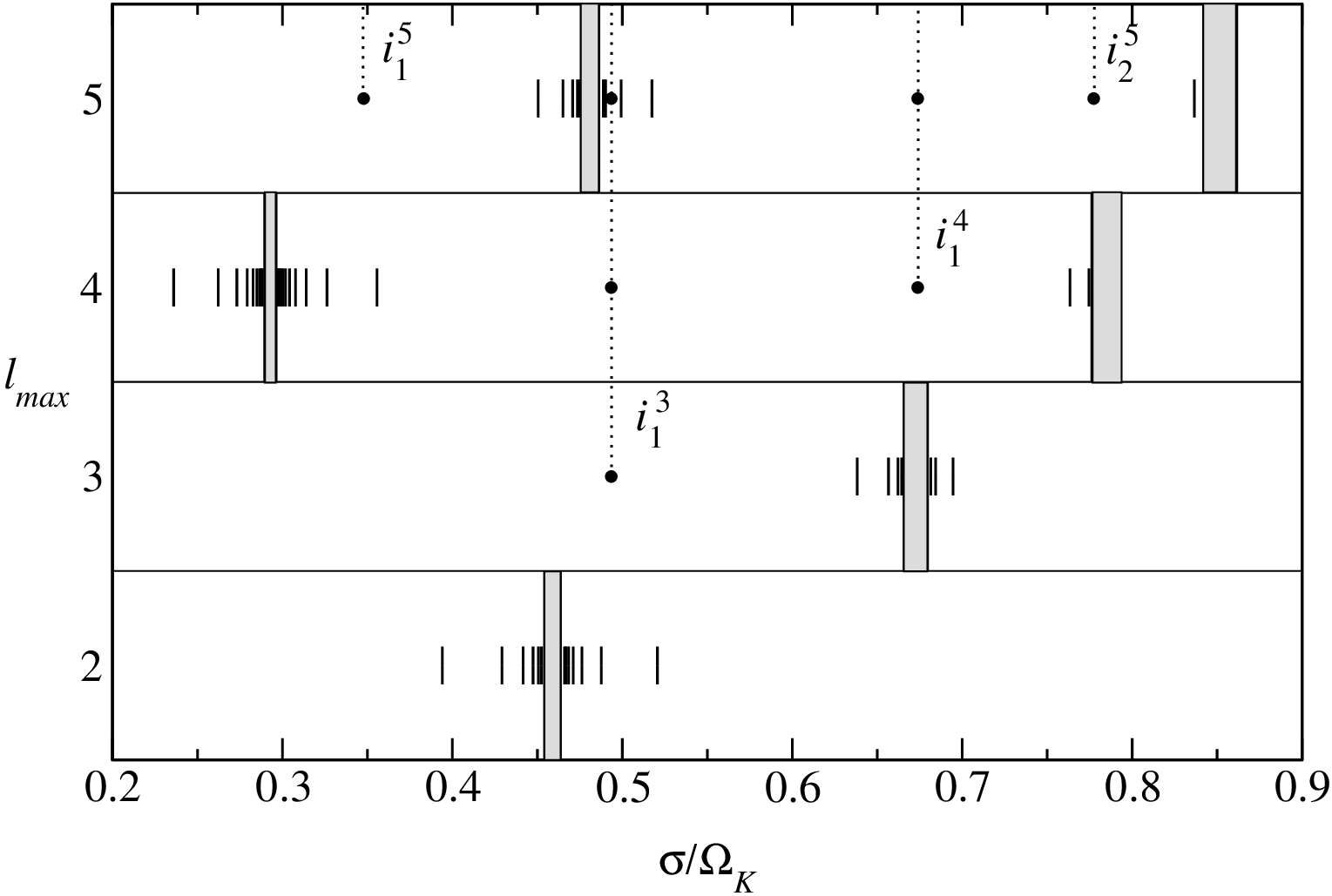}
  \caption{Modes appearing for different cut-off values $l_{\rm max}$
    for a stellar model with $M/R = 0.01$. Around each patch of the
    continuous spectrum (grey shaded areas), there are many modes,
    which dissappear as $l_{\rm max}$ is increased. The ``true''
    inertial modes, which are labeled with $i_n^l$, remain fixed as
    $l_{\rm max}$ is further increased.}
\end{figure}
\clearpage

\newpage
\begin{figure}
  \vspace*{2cm}
  \hspace*{-5mm}
  \begin{minipage}[h]{8.6cm}
    \centering
    \epsfxsize=\textwidth
    \epsfbox{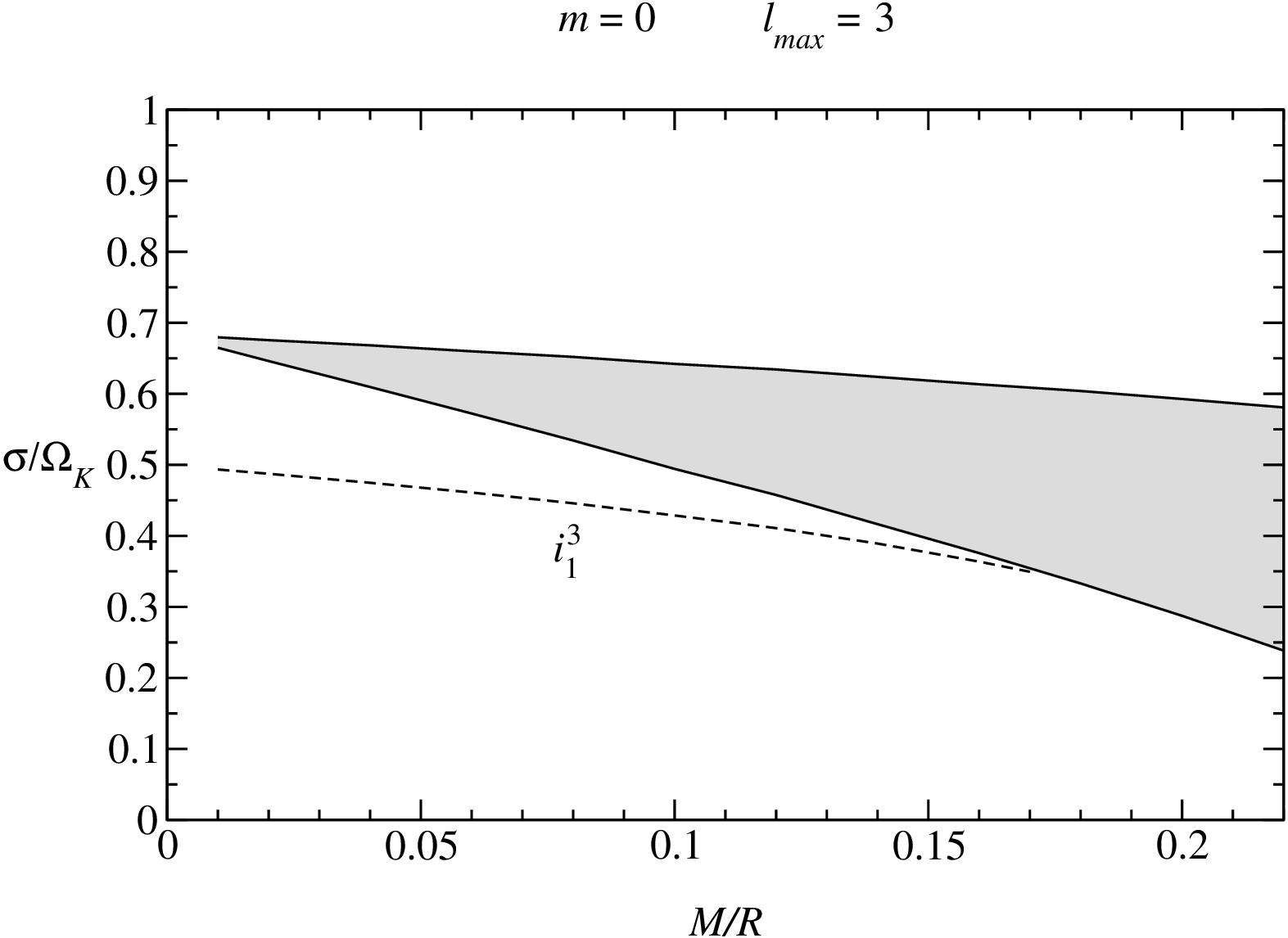}
  \end{minipage}
  \begin{minipage}[h]{8.6cm}
    \centering
    \hspace*{5mm}
    \epsfxsize=\textwidth
    \epsfbox{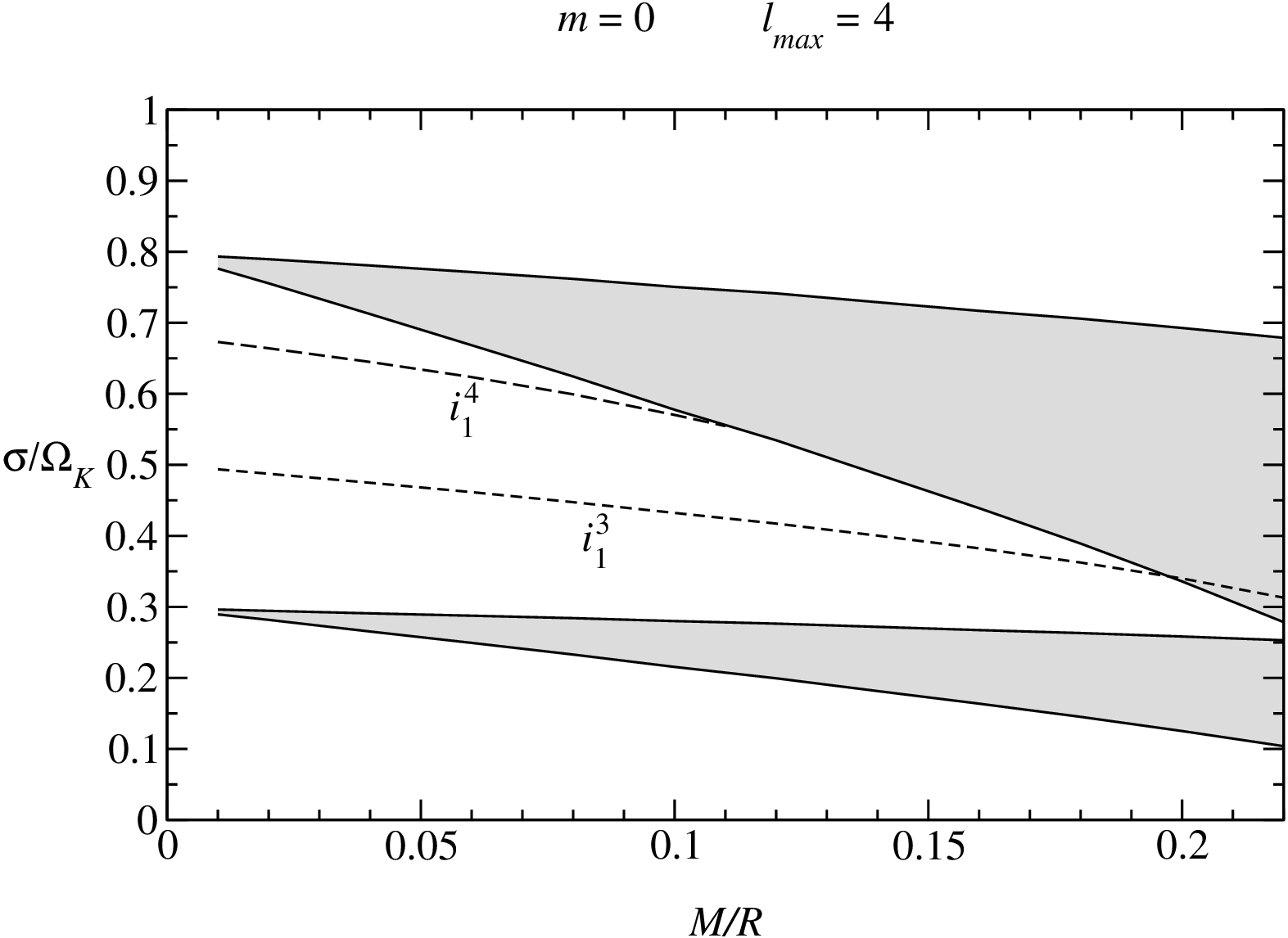}
  \end{minipage}
\end{figure}
\begin{figure}
  \hspace*{-5mm}
  \begin{minipage}[h]{8.6cm}
    \centering
    \epsfxsize=\textwidth
    \epsfbox{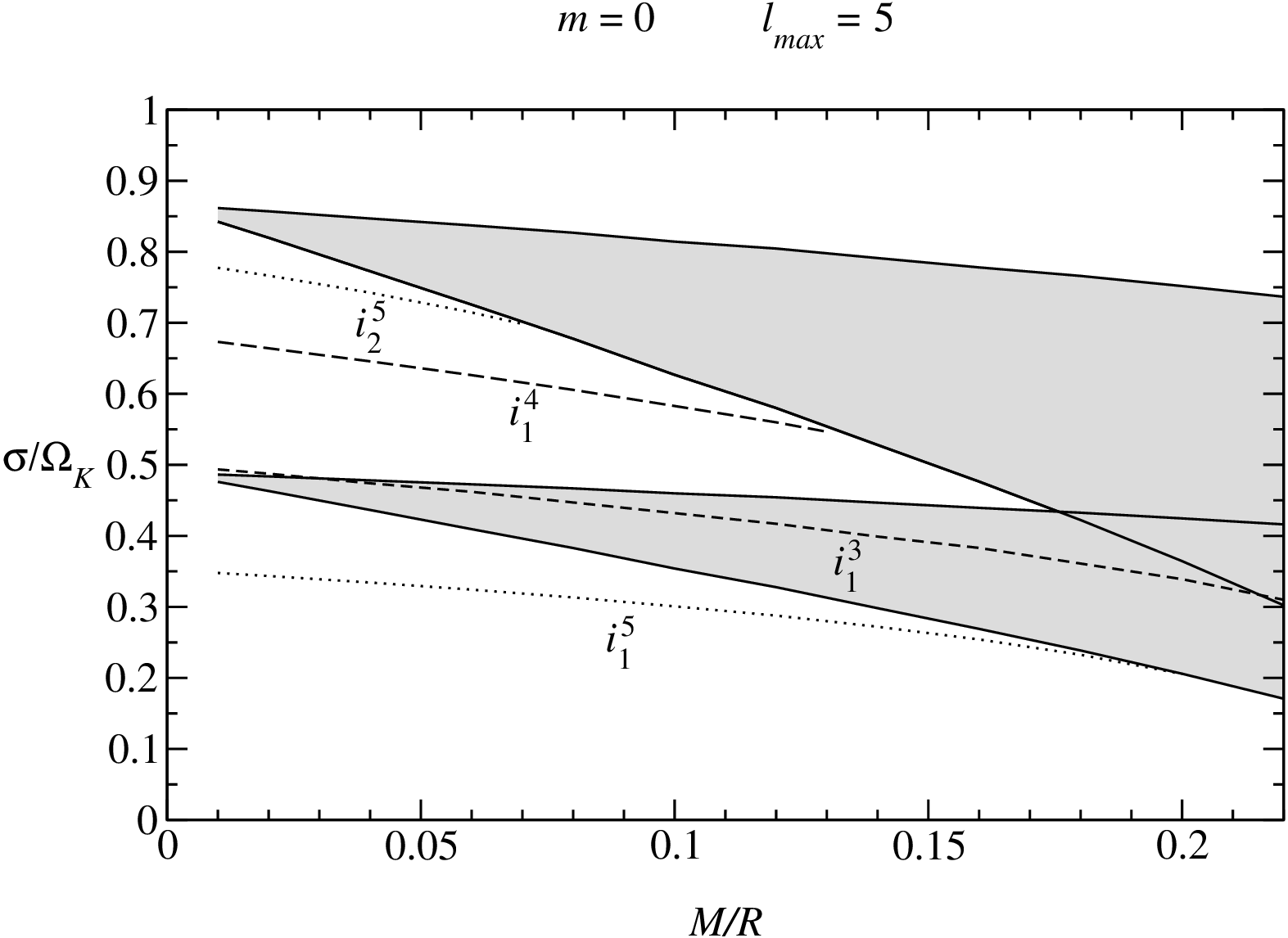}
  \end{minipage}
  \begin{minipage}[h]{8.6cm}
    \centering
    \epsfxsize=\textwidth
    \hspace*{5mm}
    \epsfbox{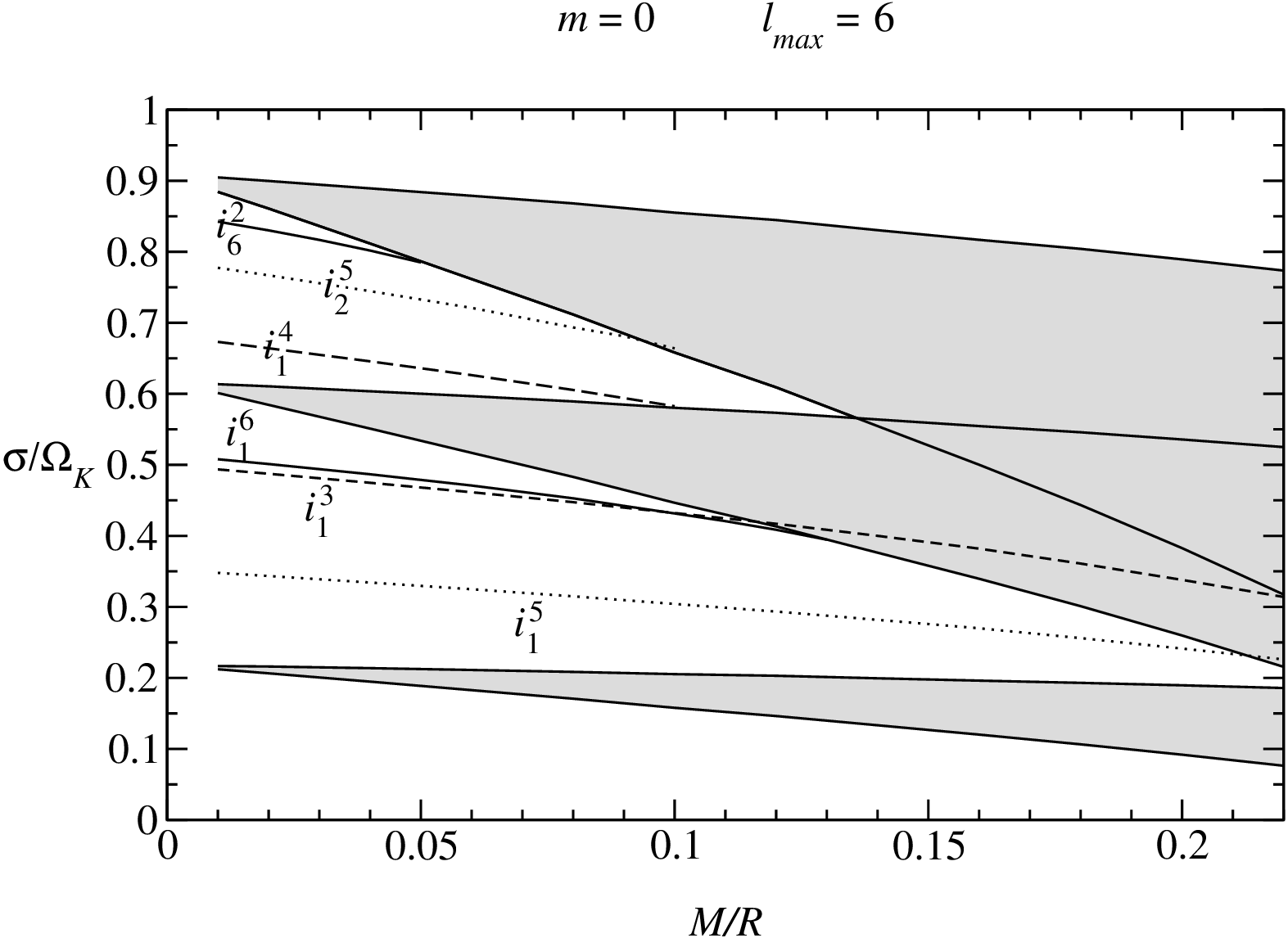}
  \end{minipage}
  \caption{The continuous spectrum and the inertial modes as functions
    of the compactness $M/R$ for $m=0$ different values of $l_{\rm
      max}$.  Only the $i_1^3$ and $i_1^5$ modes can exist for the
    whole compactness range while the other modes eventually vanish
    inside the continuous spectrum.}
\end{figure}
\clearpage

\newpage
\begin{figure}
  \vspace*{2cm}
  \hspace*{-5mm}
  \begin{minipage}[h]{8.6cm}
    \centering
    \epsfxsize=\textwidth
    \epsfbox{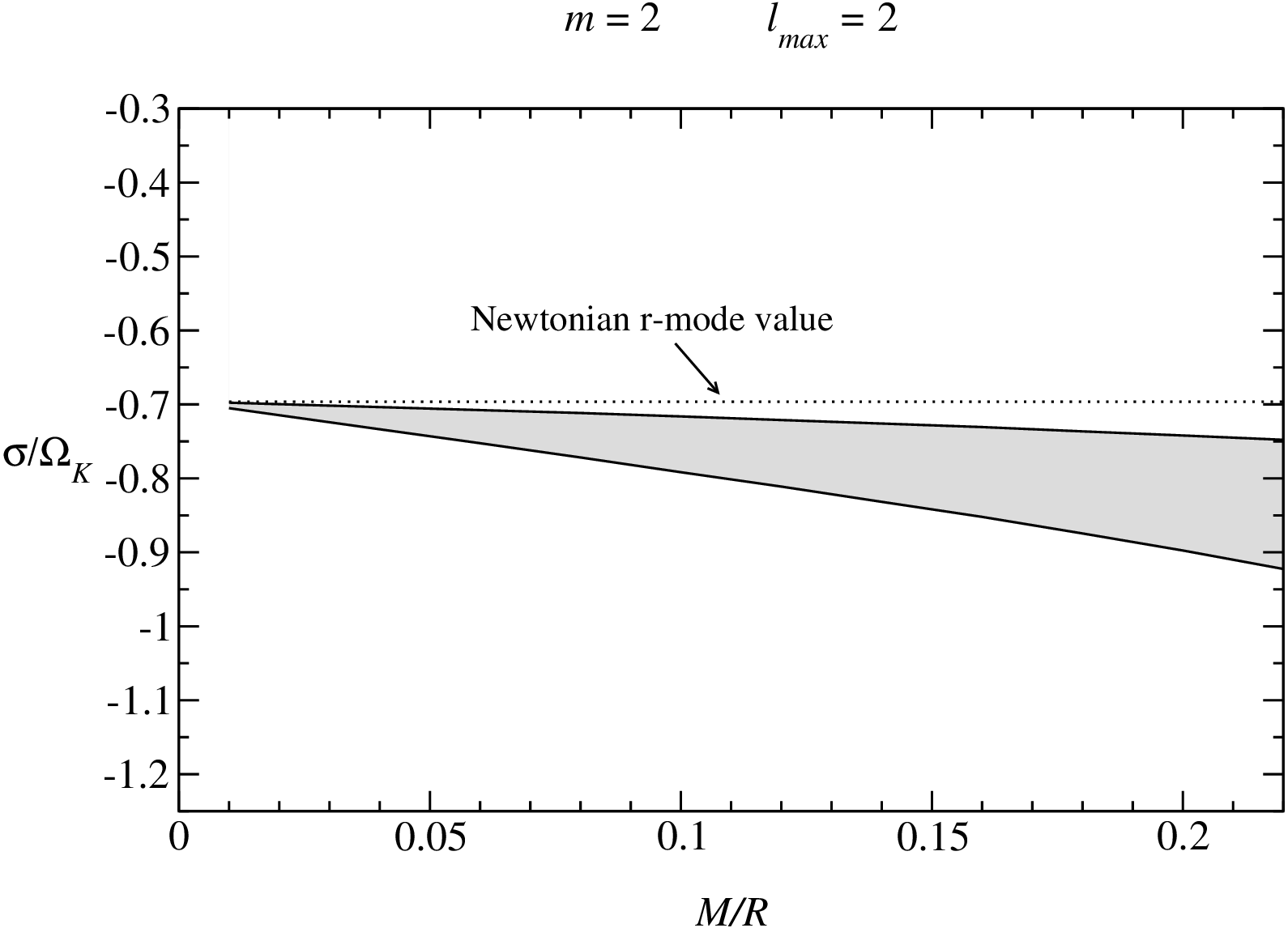}
  \end{minipage}
  \begin{minipage}[h]{8.6cm}
    \centering
    \epsfxsize=\textwidth
    \hspace*{5mm}
    \epsfbox{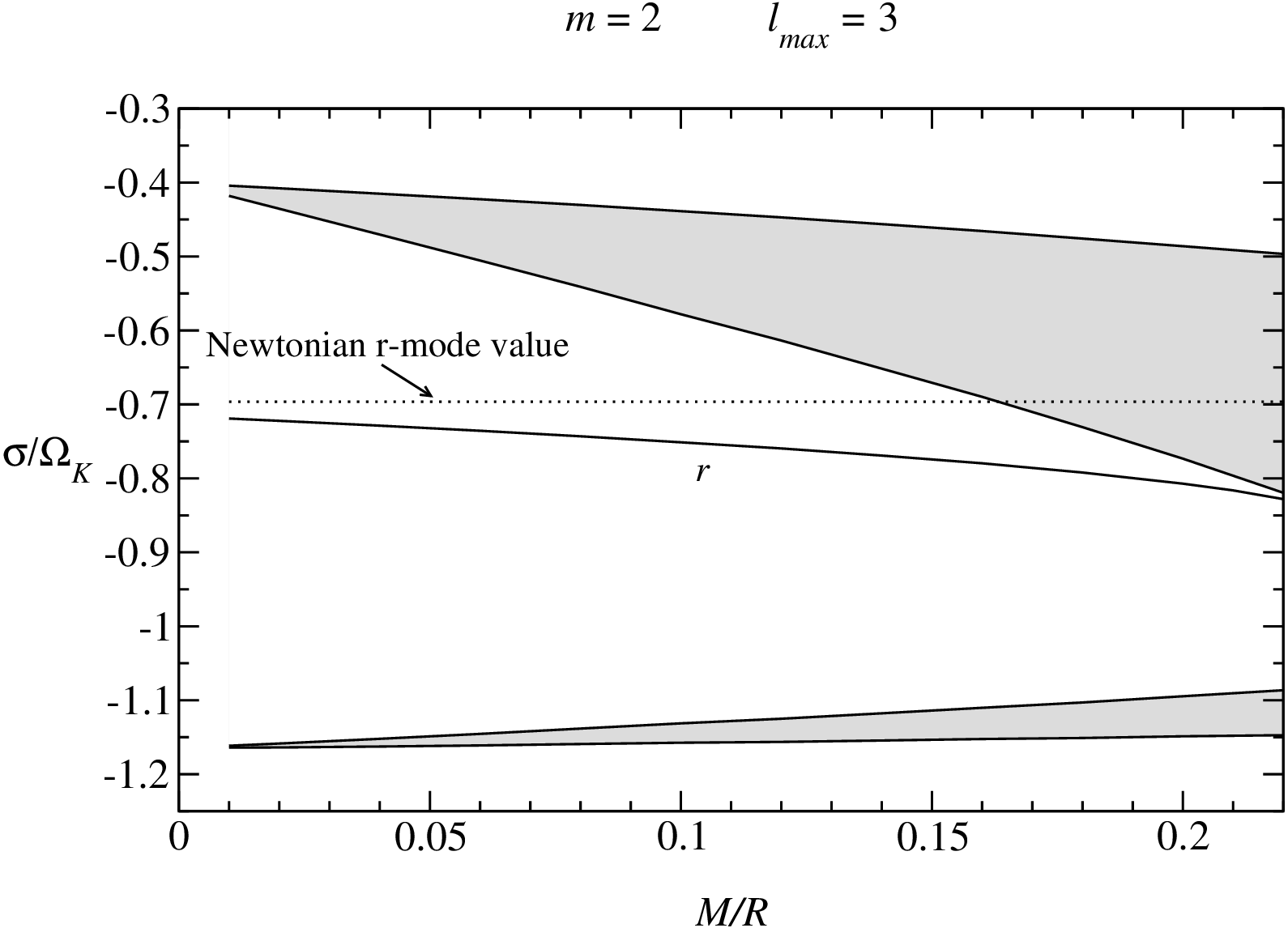}
  \end{minipage}
\end{figure}
\begin{figure}
  \hspace*{-5mm}
  \begin{minipage}[h]{8.6cm}
    \centering
    \epsfxsize=\textwidth
    \epsfbox{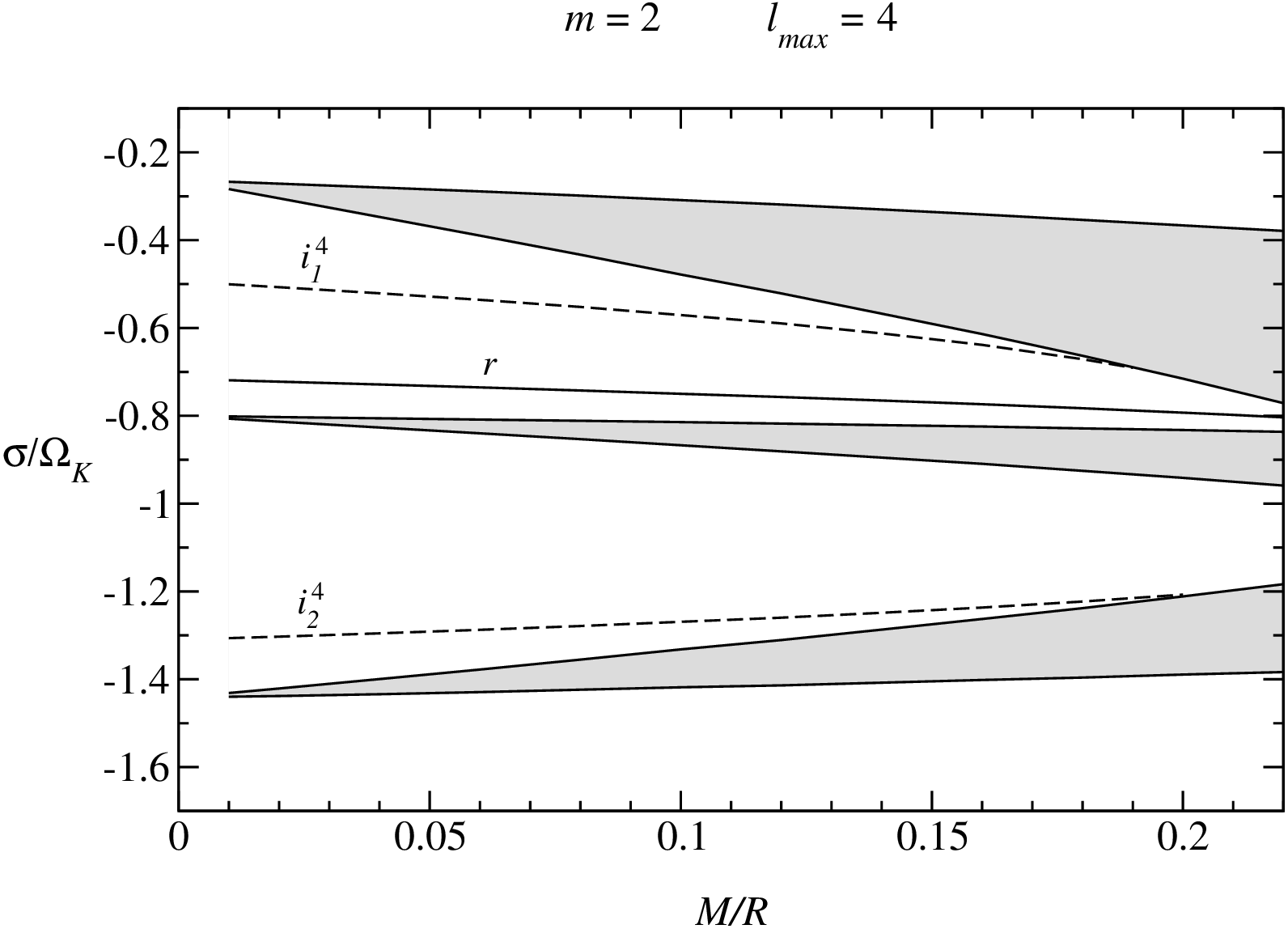}
  \end{minipage}
  \begin{minipage}[h]{8.6cm}
    \centering
    \hspace*{5mm}
    \epsfxsize=\textwidth
    \epsfbox{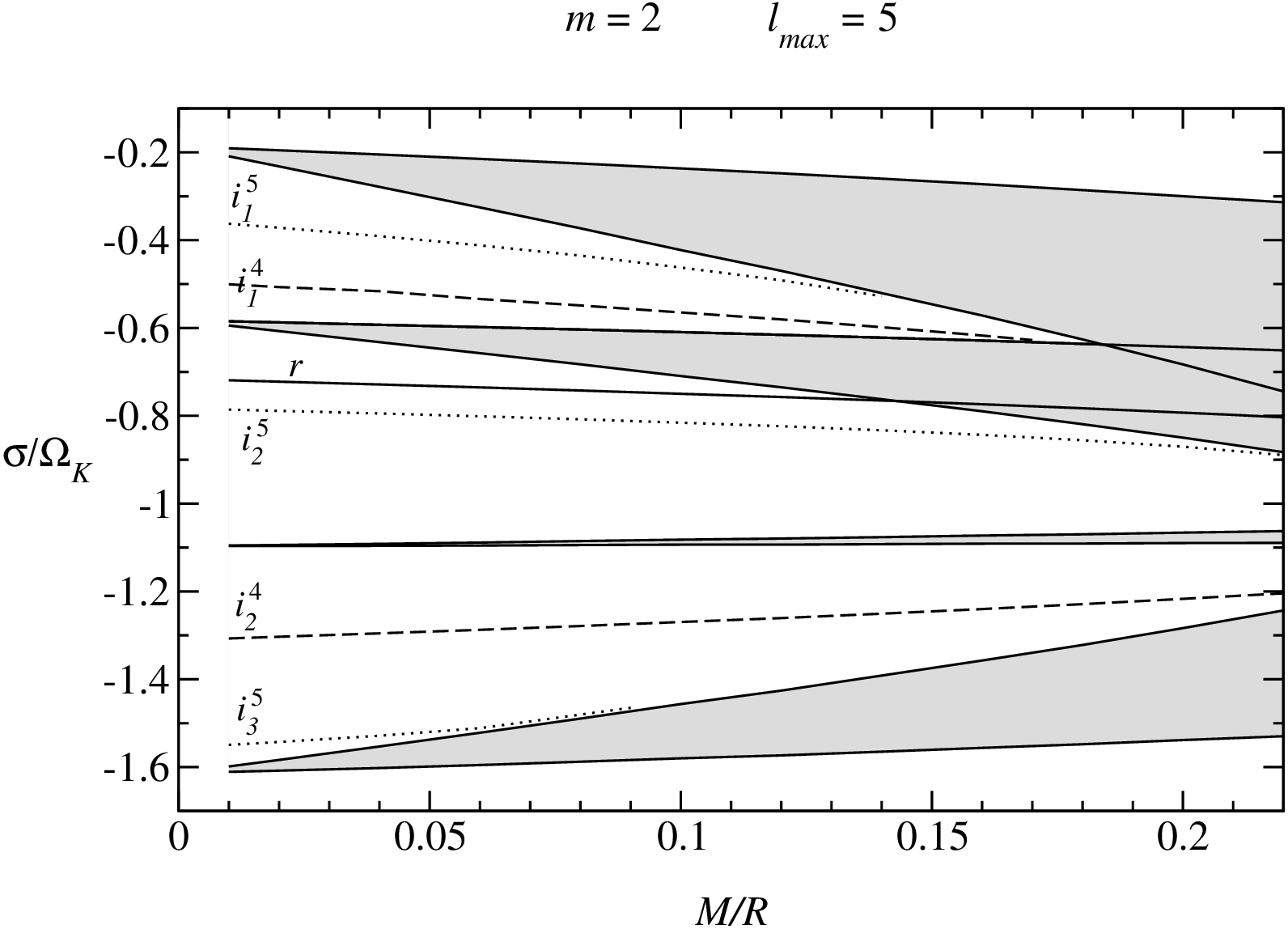}
  \end{minipage}
  \caption{Same as in Fig.~3 for $m=2$. In addition we have plotted
    the Newtonian value for the $r$-mode according to Formula (\ref{r
      mode}). The relativistic $r$ mode appears for $l_{\rm max} = 3$
    well outside the continuous spectrum as the latter has shifted
    away from its previous location for $l_{\rm max} = 2$. Only the
    $r$ mode and the $i_2^4$ and $i_2^5$ modes can exist for the whole
    compactness range.}
\end{figure}
\newpage

\begin{figure}
  \centering
  \epsfxsize=15cm
  \epsfbox{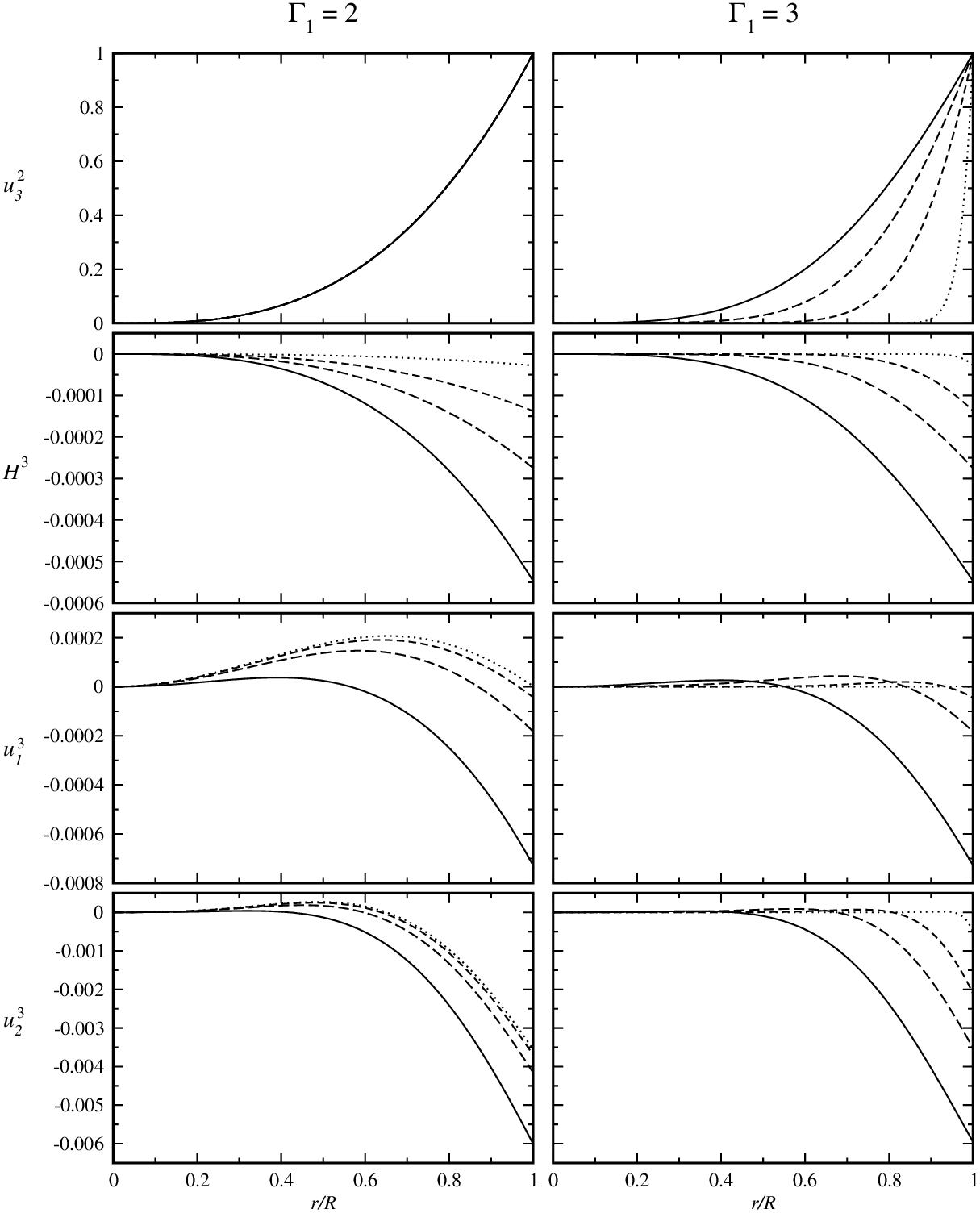}
  \caption{Sequence of $r$-mode eigenfunctions for a barotropic
    ($\Gamma_1 = \Gamma = 2$, left panels) and a non-barotropic
    ($\Gamma_1 = 3$, right panels) stellar model with $M/R = 0.01$.
    The plot style of the curves is solid for $\veps = 0.2$, long
    dashed for $\veps = 0.1$, short dashed for $\veps = 0.05$ and
    dotted for $\veps = 0.01$.}
\end{figure}


\begin{thebibliography}{99}

\bibitem{A98} Andersson N., 1998, ApJ, {\bf 502}, 708

\bibitem{AK01} Andersson N., Kokkotas, 2001, Int. J. Mod. Phys. {\bf 10} 381

\bibitem{AJK01} Andersson N., Jones D.I., Kokkotas K.D., 2001, astro-ph/0111582

\bibitem{BCL71} Battiston L., Cazzola P., Lucaroni L., 1971, Nuovo
  Cimento B, 3, 275

\bibitem{C41} Cowling T.G., 1941, MNRAS, 101, 367

\bibitem{BK99} Beyer H.R., Kokkotas K.D., 1999, MNRAS, 308, 745

\bibitem{FM98} Friedman J.L., Morsink S.M., 1998, ApJ, 502, 714

\bibitem{FSD} Font J.A., Dimmelmeier H., Gupta A., Stergioulas N.,
  2001, MNRAS, 325, 1463

\bibitem{Hartle67} Hartle J.B., 1967, ApJ, 150, 1005

\bibitem{Koj92} Kojima Y., 1992, Phys. Rev. D, 46, 4289

\bibitem{KH00} Kojima Y., Hosonuma M., 2000. Phys.Rev.D, 62, 044006

\bibitem{Koj97} Kojima Y., 1997,  Prog. Theor. Phys. Suppl., 128, 251

\bibitem{Koj98} Kojima Y., 1998, MNRAS, 293, 49

\bibitem{LAF01} Lockitch K.H., Andersson N., Friedman J.L., 2001,
  Phys. Rev. D, 63, 024019

\bibitem{LA01} Lockitch K.H., Andersson N., gr-qc/0106088

\bibitem{LF99} Lockitch K.H., Friedman J.L., 1999, ApJ, 521, 764

\bibitem{LI99} Lindblom L., Ipser J.R., 1999, Phys. Rev. D, 59, 044009

\bibitem{NumRec} Press W.H., Flannery B.P., Teukolsky S.A., Vetterling
  W.T., 1992, Numerical Recipes in C, 2nd ed. (Cambridge: Cambridge
  University Press)

\bibitem{RLS00} Rezzolla L., Lamb F.K., Shapiro S.L., 2000, ApJ Lett., 531, 139

\bibitem{RKa}Ruoff J., Kokkotas K.D., 2001, MNRAS 328, 678

\bibitem{RKb} Ruoff J., Kokkotas K.D., 2002, MNRAS 330, 1027

\bibitem{RSK02} Ruoff J., Stavridis A., Kokkotas K.D., 2002, MNRAS in press, gr-qc/0109065

\bibitem{SF01} Stergioulas N., Font J.A., 2001, Phys.Rev.Lett., 86, 1148

\bibitem{S02} Stergioulas N., 2002, private communication

\bibitem{Y01} Yoshida S., 2001, ApJ, 558, 263

\bibitem{YF01} Yoshida S., Futamase T., 2001,  Phys.Rev.D, 64, 123001

\bibitem{YL00a} Yoshida S., Lee U., 2000a, ApJ, 529, 997

\bibitem{YL00b} Yoshida S., Lee U., 2000b, ApJS, 129, 353

\bibitem{YL02} Yoshida S., Lee U., 2002, ApJ in press, gr-qc/0110038

\end{thebibliography}
\end{document}